\documentclass[conference,compsoc]{IEEEtran}
\IEEEoverridecommandlockouts
\usepackage{caption}
\usepackage{adjustbox}
\usepackage{subcaption, graphicx}
\usepackage{tabularx}
\usepackage{booktabs}
\usepackage{amsmath}
\usepackage{soul}
\usepackage{hhline}
\usepackage{subfiles}
\usepackage{tikz}
\usepackage{xcolor}
\usepackage[normalem]{ulem}
\usepackage{multirow}
\usepackage{enumitem}
\usepackage{comment}

\usepackage[labelfont=bf,format=plain,justification=raggedright,singlelinecheck=false]{caption}
\makeatletter
\newif\if@restonecol
\makeatother

\usepackage[linesnumbered,ruled,vlined]{algorithm2e}
\usepackage[utf8]{inputenc}
\usepackage{amsmath}
\usepackage{algorithmic}
\usepackage{subcaption, graphicx}
\usepackage{tabularx}
\usepackage{booktabs}
\usepackage{soul,xcolor}

\SetCommentSty{mycommfont}
\usepackage{amsthm}
\usepackage{amssymb}
\usepackage[english]{babel}
\newtheorem{definition}{Definition}

\usepackage{graphicx}
\usepackage{multirow}
\usepackage{float}
\usepackage{algorithm2e}
\usepackage{xcolor} 

\newcommand\eat[1]{}
\setstcolor{red}

\AtBeginDocument{%
  \providecommand\BibTeX{{%
    \normalfont B\kern-0.5em{\scshape i\kern-0.25em b}\kern-0.8em\TeX}}}

\begin{document}

\title{Efficient Privacy-Preserving Cross-Silo Federated Learning with Multi-Key Homomorphic Encryption}


\author{\IEEEauthorblockN{Abdullah Al Omar, Xin Yang, Euijin Choo\textsuperscript{*}, Omid Ardakanian}
\IEEEauthorblockA{
Department of Computing Science, University of Alberta
\\
\{aomar3,xyang18,euijin,oardakan\}@ualberta.ca}
\thanks{\textsuperscript{*}Corresponding author: euijin@ualberta.ca}
}

\maketitle
\begin{abstract}

Federated Learning (FL) is susceptible to privacy attacks, such as data reconstruction attacks, in which a semi-honest server or a malicious client infers information about other clients' datasets from their model updates or gradients. 
To enhance the privacy of FL, recent studies combined Multi-Key Homomorphic Encryption (MKHE) and FL, making it possible to aggregate the encrypted model updates using different keys without having to decrypt them.
Despite the privacy guarantees of MKHE, existing approaches are not well-suited for real-world deployment due to their high computation and communication overhead.
We propose MASER, an efficient MKHE-based Privacy-Preserving FL framework that combines consensus-based model pruning and slicing techniques to reduce this overhead. 
Our experimental results show that MASER is 3.03 to 8.29 times more efficient than existing MKHE-based FL approaches in terms of computation and communication overhead while maintaining comparable classification accuracy to standard FL algorithms. Compared to a vanilla FL algorithm, the overhead of MASER is only 1.48 to 5 times higher,
striking a good balance between privacy, accuracy, and efficiency in both IID and non-IID settings.
\end{abstract}


\section{Introduction}
Cross-silo Federated Learning (FL) has emerged as a viable decentralized machine learning approach enabling multiple institutions, referred to as clients, to collaboratively train a model without sharing their private data~\cite{kairouz2021advances}. 
In FL, a central server aggregates model parameter updates made independently by clients to produce a global model, which is then sent to clients for further training with their local data. 
This decentralized approach is particularly valuable in sectors such as healthcare, public safety, and finance, where high-quality models are needed, but data privacy is paramount. \looseness=-1

Although FL seemingly protects data privacy by eliminating data sharing with the central server, it has been shown that the model updates or gradients can still leak private information~\cite{zhu2019deep, yin2021see, geiping2020inverting}. In recent years, various privacy-preserving techniques, such as Differential Privacy (DP)~\cite{wei2020federated}, Secure Multi-Party Computation (SMPC)~\cite{bonawitz2017practical}, and Homomorphic Encryption (HE)~\cite{batchcrypt, jin2023fedml}, have been proposed to mitigate the privacy risks in FL. 

DP protects privacy by introducing calibrated noise into the model updates, but this can degrade the model performance, particularly in complex learning tasks~\cite{tramer2021differentially}.
SMPC offers strong privacy guarantees by providing a zero-knowledge framework for aggregating model parameter updates, but this comes at the expense of significant overhead and complex protocols~\cite{yang2019federated}. 
Specifically, clients need to repeatedly exchange intermediate results to perform secure operations, leading to significant computation and communication overhead~\cite{bonawitz2017practical}.  
These limitations make SMPC less practical for real-world applications, especially in cross-silo FL where efficient communication is critical. 

Homomorphic Encryption (HE), on the other hand, provides strong privacy guarantees without introducing noise (like DP) or requiring complex, multi-party protocols (like SMPC). 
The key feature of HE is the homomorphism property~\cite{rivest1978data} 
which enables the computation to be performed directly on encrypted data without the need to decrypt it first. 
In FL, this allows the central server to aggregate encrypted client updates, 
protecting the confidentiality of individual client data. 
This makes HE particularly well-suited for cross-silo FL scenarios, 
where institutions holding large datasets require highly accurate models 
but do not fully trust the central server not to perform intrusive inferences, e.g., to recover the training samples.

Several HE-based FL approaches have been proposed, most of which employ single-key HE schemes in which all clients share the same public and secret key pair~\cite{batchcrypt,jin2023fedml,fedphe,maskcrypt}. 
This poses a risk, as a malicious client can decrypt other clients' updates 
using the shared key and perform intrusive inferences on their data. 
To tackle this issue, efforts have been made recently to adopt Multi-Key HE (MKHE)-based schemes, 
where each client uses its own key pair for encryption and decryption~\cite{10334062, ma2022privacy, park2022privacy}, in FL. 
This provides stronger privacy protection in that even if one client is malicious, the privacy of the other clients remains intact.



However, HE-based FL approaches introduce significant overheads due to computationally expensive cryptographic operations and increased size of encrypted data to be transmitted~\cite{jin2023fedml,fedphe}. 
To address the high overhead of HE-based FL while maintaining its privacy guarantees, several optimization techniques have been proposed, such as packing~\cite{fedphe}, batching~\cite{batchcrypt}, and masking~\cite{maskcrypt}. 
Despite these efforts, the packing and batching techniques that compress and aggregate updates 
continue to face high computational cost and memory demand, particularly for large models where encryption costs scale with model size~\cite{jin2023fedml, fedphe}. Masking techniques that prune unimportant parameters often degrade performance in non-IID settings, as they fail to account for data heterogeneity across clients~\cite{maskcrypt, li2020federated}




To address these challenges, we propose MASER, an efficient MKHE-based privacy-preserving FL framework that guarantees strong privacy while significantly reducing the overheads of HE and maintaining model performance even in the presence of data heterogeneity. The key idea is to effectively sparsify the model prior to encryption, significantly reducing computation and communication overhead of MKHE-based FL, while preserving model performance.   

MASER employs magnitude-based weight pruning to
identify a sparse subnetwork consisting of critical weights from models trained by each client, and incorporate a consensus algorithm to ensure agreement among clients on how to sparsify their models before encryption. Specifically, clients compute the L1-norm of model parameters~\cite{Han2015} after each round of local training and rank them accordingly. Then, the most important parameters are selected based on a threshold, e.g., 10\% of the top-ranked weights. 
Each client generates a mask that indicates the selected (important) parameters for encryption.
The \textit{local masks} generated by clients are not necessarily the same, as the important parameters are chosen based on their local dataset. 
To reach a consensus among clients on what parameters should be retained and encrypted, 
the local masks are sent to the central server that uses a majority voting strategy to generate a \textit{global mask}. 
This majority voting approach offers robustness against malicious clients generating \textit{poisoned ranks}~\cite{FRL}.
Upon receiving the global mask, each client applies this mask to its trained model to sparsify the model. 
To further minimize the computation and communication overhead, we incorporate slicing by dividing the parameters of the resulting sparse model into smaller chunks tailored to the MKHE module's key size (e.g., 8192). The sliced parameters are encrypted and then transmitted to the server for aggregation. By reducing the need for full model encryption and focusing on the most important model parameters, 
MASER achieves significant efficiency gains without sacrificing accuracy or privacy, 
making it well-suited for real-world cross-silo FL applications.\looseness=-1

\noindent\textbf{Contribution:} Our contribution is threefold:
\begin{itemize}
    \item We present a practical privacy-preserving cross-silo FL framework utilizing MKHE to effectively counter privacy attacks from an honest-but-curious server and malicious clients.
    \item We propose MASER, a novel strategy that combines a consensus-based sparsification method to find and update the important parameters and a selective homomorphic encryption scheme to encrypt these parameters, ensuring that sparsification has negligible impact on model performance in MKHE-based FL while significantly reducing the overhead introduced by MKHE. To our knowledge, this is the first work on privacy-preserving FL that achieves substantial reductions in the encrypted model size while maintaining model performance and privacy guarantees. 
    \item We evaluate our system across multiple datasets and neural network architectures, in IID and non-IID settings,
    demonstrating its performance, privacy protection capability, and robustness in real-world scenarios. Our approach requires only $1.48{-}5{\times}$ more time than vanilla FL to deliver a privacy-preserving solution, without compromising model accuracy. In comparison, the traditional MKHE-based FL algorithm 
    incurs significantly higher overhead, requiring approximately $4.7{-}39{\times}$ more time than vanilla FL. This highlights the efficiency and practicality of MASER for privacy-preserving federated learning.
\end{itemize}

\noindent\textbf{Organization of the paper:} The rest of this paper is organized as follows. Section 2 presents background information on FL, MKHE, and neural network pruning. Section 3 describes our threat model. Section 4 describes our methodology, detailing the MASER framework and its key components, namely magnitude-based pruning, slicing, and majority voting. Section 5 provides an in-depth experimental analysis, evaluating the performance of MASER under real-world conditions. Section 6 discusses related work. Finally, Section 7 concludes the paper.


\section{Background}
This section introduces the key concepts underlying our approach, MASER, including federated learning, multi-key homomorphic encryption, and neural network weight pruning.

\subsection{Federated Learning}
Cross-silo FL is a machine learning approach that offloads model training from a central server to multiple clients (i.e., institutions) that hold their own data and wish to participate in the model training task~\cite{huang2022cross}.
By training the model on decentralized data, 
FL forgoes sharing private client data with a central server. 
Instead, the clients participating in each round of training 
train the model locally using their own data and share the \textit{local model} update with a central aggregation server 
to update a \textit{global model} that can generalize well to all clients. 
The updated global model is then sent to clients and used in the next round.

Consider a neural network model $\hat{y} = f(x; \theta)$, where $x$ is the input data, $\theta$ represents the model parameters, and $\hat{y}$ is the predicted label for the input $x$.
To train this neural network, we minimize the loss, a function of $\theta$ and a measure of the difference between $\hat{y}$ and the ground truth label $y$, denoted as $l(y_i, \hat{y}_i)$.
The loss across samples in a dataset $\mathcal{D}$ is given by:
\begin{equation}
    \mathcal{L}(\theta; \mathcal{D}) = \frac{1}{|\mathcal{D}|} \sum_{(x_i, y_i) \in \mathcal{D}} l(y_i, \hat{y}_i),
\end{equation}
where $|\mathcal{D}|$ is the total number of data samples, $i$ is the index of data sample, $y_i$ is the ground truth label for input data $x_i$, and $\hat{y}_i$ is the predicted label. 
In gradient-based optimization algorithms, such as Stochastic Gradient Descent~(SGD)~\cite{amari1993backpropagation}, the model parameters are updated iteratively as follows:
\begin{equation}
\theta^{t+1} = \theta^t - \eta \cdot g^t,
\end{equation}
where $t$ is the index for the current training iteration, $\eta$ is the learning rate, and $g^t$ is the gradient of loss with respect to the parameters.


In FL, the training process starts with the aggregation server initializing the machine learning model $\theta^0$ and distributing the model to all or a subset of participating clients.
Each client $i$ updates the received model by training it on its local dataset $\mathcal{D}_i$.
Thus, the local update of client $i$ at global training round $t$ is governed by: 
\begin{equation}
    g^t_i = \nabla_{\theta_i} \mathcal{L}(\theta_i^t; \mathcal{D}_i).
\end{equation}
Once the local training completes,
each client sends its local model parameters $\theta^t_i$ to the central aggregation server. The server then aggregates the local models, for example using the Federated Averaging (FedAvg)  algorithm~\cite{mcmahan2017communication}:
\begin{equation}
    \theta^{t+1}= \sum_{i=1}^N \alpha_i \theta^t_i,
\end{equation}
where $\alpha_i = \frac{|D_i|}{|D|}$ is the weight assigned to client $i$ based on the size of its local dataset and $N$ is the total number of clients participating in the model training in that round.
The aggregated model $\theta^{t+1}$ will be distributed to the participating clients for the next round of training until convergence.

FedAvg is effective when the data held by each client is Independent and Identically Distributed (IID). 
However, the effectiveness of FedAvg deteriorates drastically when the data is non-IID \cite{zhao2018federated}. 
To address this problem and tackle data heterogeneity in federated learning, 
Li~\textit{et al.} proposed FedProx~\cite{li2020federated}, which improves the effectiveness of FL by
adding a proximal term to local subproblems:
\begin{equation}
\label{eq:fedprox}
    \theta_i^{t} = \arg\min_{\theta} \left\{\mathcal{L}(\theta; \mathcal{D}_i) + \frac{\mu}{2}\|\theta-\theta^{t-1}\|^2 \right\},
\end{equation}
where the proximal term $\frac{\mu}{2}\|\theta-\theta^{t-1}\|^2$ restricts
the local updates to be closer to the global model and eliminates the need 
to manually set the number of local epochs.

\subsection{Multi-key Homomorphic Encryption} 
\label{subsec:bg_mkhe}
Homomorphic encryption 
enables encrypting plaintext data and performing computation (usually simple arithmetic operations) on the encrypted ciphertext without having to first decrypt it, such that the result after decryption is consistent with the computation on the plaintext data.
In the context of FL, the encrypted data are model parameters and the computation performed on encrypted data is model aggregation.
The MKHE scheme lets clients encrypt their model updates with different keys. 
The server can still aggregate them without accessing the decryption keys, thus enhancing privacy.

\begin{definition}[RLWE~\cite{lyubashevsky2013ideal}]
Let $\mathcal{R}_q$ be a polynomial ring and
$\psi$ be the error distribution over $\mathcal{R}_q$ with $q$ being a prime integer.
Given a secret polynomial $s(x)$ chosen from the dual fractional ideal of $\mathcal{R}_q$,
we generate a sample $(a_i(x), s(x){\cdot}a_i(x){+} e_i(x))$ by
choosing $a_i(x)$ from $\mathcal{R}_q$ uniformly at random and sampling $e_i(x)$ from $\psi$.
The problem of Ring Learning with Errors (RLWE) concerns distinguishing arbitrarily many independent pairs of the form
$$
\Big(a_i(x),b_i(x)\Big) = 
\Big(a_i(x),s(x){\cdot}a_i(x) + e_i(x)\Big)\in \mathcal{R}_q^2$$
from uniformly random and independent pairs.
\end{definition}
Let $\mathcal{R}=\mathbb{Z}[x]/(x\textsuperscript{n}+1)$ be the cyclotomic ring with a power-of-two dimension $n$, where $\mathbb{Z}[x]$ is a polynomial ring with integer coefficients,
and $\mathcal{R}_q = \mathcal{R}/\langle q\rangle$. 
For this ring and the appropriate choice of error distribution,
the RLWE problem is hard~\cite{lyubashevsky2013ideal}.
Following the previous work~\cite{maskcrypt,ma2022privacy}, 
we use CKKS, and in particular MKCKKS
which is a multi-key variant of CKKS~\cite{chen2019efficient},
as our homomorphic encryption algorithm. 
MKCKKS enables clients to independently encrypt their updates, thereby enhancing privacy by keeping each client's data confidential. Its compatibility with real-number arithmetic is ideal for FL, where numerical precision is critical for model updates. 
xMKCKKS~\cite{ma2022privacy} is an extension of MKCKKS that uses a common public key for encryption rather than the public key generated by each client.
This common public key is the sum of the public keys of all clients.
By using public key aggregation, xMKCKKS minimizes privacy risks in the decryption phase, ensuring that decryption accuracy and privacy are maintained even with multiple client keys. 
This makes xMKCKKS suitable for cross-silo, privacy-preserving FL.
We adopt the public key aggregation method proposed in xMKCKKS in MASER.

Let us denote the security parameter by $\lambda$, the secret key distribution by $\mathcal{X}$, and the space of local models by $\mathcal{M}$.
The functions of MKHE are defined below:
\begin{itemize}
 \item \texttt{Setup}$(1^\lambda)$: Given the security parameter $\lambda$ for this scheme, this function defines the \textit{public parameters (pp)}=$(n,q,\mathcal{X},\psi, a)$. These public parameters are the same for all the clients. 
 \item \texttt{KeyGen}$(pp)$: This function is invoked by each client $i$ to generate their  public key and secret key. 
 Specifically, the secret key is sampled from $\mathcal{X}$ 
 and the error vector is sampled from $\psi^d$. 
 Here, $d$ denotes the dimension of the error vector, aligning it with the structure of the polynomial ring \( \mathcal{R}_q^d \) 
 from which $a$ was sampled uniformly at random in \texttt{Setup}.
 The public key, denoted $pk_i$, is then calculated using the corresponding secret key $sk_i$, the error vector $e_i$, and the random vector $a$. Specifically $pk_i = -sk_i \cdot a + e_i$ mod $q$ in $\mathcal{R}^d_q$.  
 Clients will share their $pk_i$ with the key manager for aggregation. To aggregate public keys, we use the same scheme as in xMKCKKS~\cite{ma2022privacy}:
 \begin{equation}
      pk= \sum_{i=1}^N pk_i= \sum_{i=1}^N (-sk_i) \cdot a + \sum_{i=1}^N e_i \pmod q.
 \end{equation}
The aggregated public key $pk$ prevents an honest-but-curious server from directly decrypting the ciphertext shared by each client. It is distributed to all clients for homomorphic encryption. 
\item \texttt{Enc}$(\theta_i;pk)$: Let the model parameters of client $i$ be $\theta_i \in \mathcal{R}$. For encryption, we use the aggregated public key $pk$ along with random vectors $a$ and $b$, where $a = a[0]$ and $b = b[0]$. Here, $a[0]$ and $b[0]$ indicate that we are taking the first elements of these vectors to simplify and optimize the encryption process. Additionally, errors $e_0$ and $e_1$ are sampled from the error distribution $\psi$. This function returns the encrypted weight parameters $\theta_i = (d_0, d_1) \in \mathcal{R}^2_q$. The values $d_0$ and $d_1$ are used to sample the weight parameters in the ring $\mathcal{R}^2_q$, where $d_0 = pk \cdot b + \theta_i + e_0 \pmod{q}$ and $d_1 = pk \cdot a + e_1 \mod{q}$. Here, $d_0$ is generated using the actual update $\theta_i$ of client $i$, while $d_1$ is generated with only the error component.

 A level-$l$ multi-key encryption of a client's model parameters $\theta_i$ with respect to the secret keys $sk_i= (1,sk_1, sk_2, sk_3, ..., sk_i)$ is a vector in the ciphertext space $\theta_i =(d_0, d_1, d_2, d_3, ..., d_i) \in \mathcal{R}^{(k+1)}_{ql}$ satisfying $\langle \theta_i,sk_i\rangle \approx \theta_i \pmod q_l$. In the case of a homomorphic operation, e.g., the addition of $\theta_i$ and $\theta_j$, which are two ciphertext vectors, this operation returns an encrypted $\theta_k$ such that $\langle \theta_k,sk_i\rangle_{ql}$ is approximately equal to $\theta_i+\theta_j$.
\item \texttt{PartialDec}$(\theta_i, sk_i)$: In partial decryption, each client $i$ will have the secret key $sk_i\in\mathcal{R}$. Given a polynomial $d_i$ and a secret key $sk_i$, they will sample an error $e_i\gets \psi$ for noise flooding and return $p_i= d_1\cdot sk_i + e_i \mod q$.
\item \texttt{Merge}$(ct_0,\{p_i\}_{1\leq i \leq k})$: This function outputs $p = ct_0 + \sum_{i=1}^k p_i \mod q$, where $ct$ denotes the encrypted ciphertext. 
For a multi-key ciphertext at global round \( t+1 \), $\theta^{t+1} = (ct_0, ct_1, \dots, ct_k)$, 
MKHE performs the merge phase by calculating 
\[
p = ct_0 + \sum_{i=1}^k p_i = ct_0 + \sum_{i=1}^k e_i \approx \langle \theta^{t+1}, sk_i \rangle \pmod q.
\]
Here, \( ct_0 \) serves as the initial component of the encrypted model update, and each \( p_i \) is a partial decryption contributed by a client. The merge process approximates the inner product of the model update $\theta^{t+1}$ with the secret key $sk_i$. After merging, scaling factors are removed to retrieve the plaintext model update $\theta^{t+1}$.

\end{itemize}
\subsection{Neural Network Pruning}
\label{subsec:bg_pruning}

Neural network pruning techniques can be broadly classified into two categories according to when pruning is carried out. 
The first category includes techniques that assign a score to every parameter in a neural network \textit{after training} and then remove parameters with the lowest scores.
Numerous score functions have been defined for pruning after training, such as the weight magnitude which is used in magnitude-based pruning (MP)~\cite{Han2015},
the Hessian of the training loss~\cite{Hassibi92,LeCun89},
and lookahead distortion~\cite{park2020lookahead}.
MP removes weights with the smallest magnitudes given that they contribute less to the model's output and can be pruned without significantly affecting its performance. 
Specifically, given a pruning threshold $\kappa$, 
a mask $M$ whose entries are either 0 or 1
can be generated as follows:
\begin{equation}
    M_{(j,k)} =
\begin{cases}
1, & \text{if } |W_{(j,k)}| \geq \kappa, \\
0, & \text{otherwise}, 
\end{cases}
\end{equation}
where $W_{(j,k)}$ represents the $k$-th element of the model weight vector at the $j$-th layer.
The mask is applied to the weights of the neural network by performing element-wise multiplication.
Note that in magnitude-based pruning, biases in the model parameters are not pruned.
It can be shown that MP is actually minimizing the worst-case distortion in the output of a neural network~\cite{park2020lookahead}.
MASER adopts MP for its simplicity and remarkable performance in terms of the sparsity-accuracy trade-off~\cite{Gale19}.


The second category includes techniques that prune a neural network \textit{at initialization}.
In particular, they use heuristics to identify and remove unimportant connections in a randomly initialized neural network without training it.
Examples of these techniques are Single-shot Network Pruning~(SNIP)~\cite{lee2018snip}, Gradient Signal Preservation~(GraSP)~\cite{wang2020picking}, and
Iterative Synaptic Flow Pruning (SynFlow)~\cite{tanaka2020pruning}.
Although these techniques can significantly reduce the pruning cost by eliminating the need for model training,
the pruned network does not necessarily achieve the same accuracy as the network pruned using the conventional MP technique, because the heuristics may not always be good~\cite{cheng2024surveypruning}.
Since model training happens in multiple rounds in FL, we are not restricted to single-shot foresight pruning techniques and can use MP to find a better sparse subnetwork.
In Section~\ref{sec:baseline}, we implement a baseline that uses GraSP for model pruning, as a representative of pruning at initialization techniques, and compare it with MASER, which uses MP for pruning after training. \looseness=-1
\noindent \textbf{Consensus-Based Model Pruning in Federated Learning.} Each client in FL is responsible for pruning its own model by generating a mask, 
and this task cannot be delegated to the server for privacy reasons. The masks generated by different clients may differ significantly, 
especially in the presence of data heterogeneity (the non-IID setting).
Aggregating models sparsified by different masks would hamper the convergence of FL.
Consensus-based model pruning~\cite{maskcrypt, babakniya2023revisiting, gez2023masked} addresses this challenge by reaching a consensus on the mask that will be applied for all clients to align the clients' model updates so that these updates can be aggregated in a meaningful manner.
Majority voting is an efficient consensus mechanism as it can reach a consensus among all clients using only one additional communication round and no data replication. 
Note that for pruning at initialization, clients need to reach a consensus on the mask only once before FL training, whereas for pruning after training, a new consensus needs to be reached in every training round.

\section{Threat Model}
We consider the aggregation server an honest-but-curious~(HBC) adversary, which is a common assumption in the literature~\cite{aono2017privacy,3133982}. This passive adversary performs model aggregation faithfully and adheres to the protocol. However, it is curious to extract private information from the data shared by the clients, such as model updates in FL. 


We assume that clients may be malicious; however, the total number of colluding malicious clients is not enough to compromise the consensus mechanism that we used, i.e., majority voting\footnote{In cases where the assumption on the limited number of colluding malicious clients is relaxed, more complex consensus protocols, such as those designed to handle Byzantine faults, would be necessary instead of majority voting.}. Malicious clients are not obligated to follow the protocol and may engage in adversarial actions, such as 
submitting bogus masks to the aggregation server.
This model considers the highest level of threat from clients, requiring robust mechanisms to mitigate the impact of these adversarial actions to ensure the integrity of the system and data privacy.

It is important to note that MASER is specifically designed for cross-silo FL, where clients are organizations with potentially conflicting objectives. In such settings, clients may be motivated to infer information about the training data held by their competitors, which intensifies the need for privacy-preserving mechanisms. 
MASER addresses this by providing security measures that ensure client data privacy even in the presence of curious and potentially adversarial organizations. \looseness=-1

\section{Methodology}
In this section, we introduce MASER (MASking at Each Round), a practical privacy-preserving cross-silo FL framework that is robust to an HBC aggregation server and malicious clients by leveraging MKHE. MASER is designed to address the high overhead of HE-based FL while achieving performance comparable to vanilla FL, which is vulnerable to privacy attacks.
To this end, it incorporates a pruning technique to reduce the number of model parameters that must be encrypted and sent to the server, while minimizing the impact on model performance. 
It also incorporates a majority voting strategy so that clients can reach a consensus about the most important parameters that must be retained in the reduced model, allowing model aggregation to be performed correctly.


We begin with an overview of MASER in Section~\ref{sec:overview}. Next, we describe our model training and sparsification technique that reduces the overhead of MKHE-based FL in Section~\ref{sec:sparsification}. In Section~\ref{sec:slicing}, we introduce the parameter slicing technique for efficient homomorphic encryption and model aggregation. Finally, we describe the decryption and model reconstruction process 
in Section~\ref{sec:reconstruction}. 

\begin{figure}[!t]
    \centering
    \includegraphics[clip, trim= 0.7cm 0.6cm 0.7cm 0.6cm, width=\linewidth]{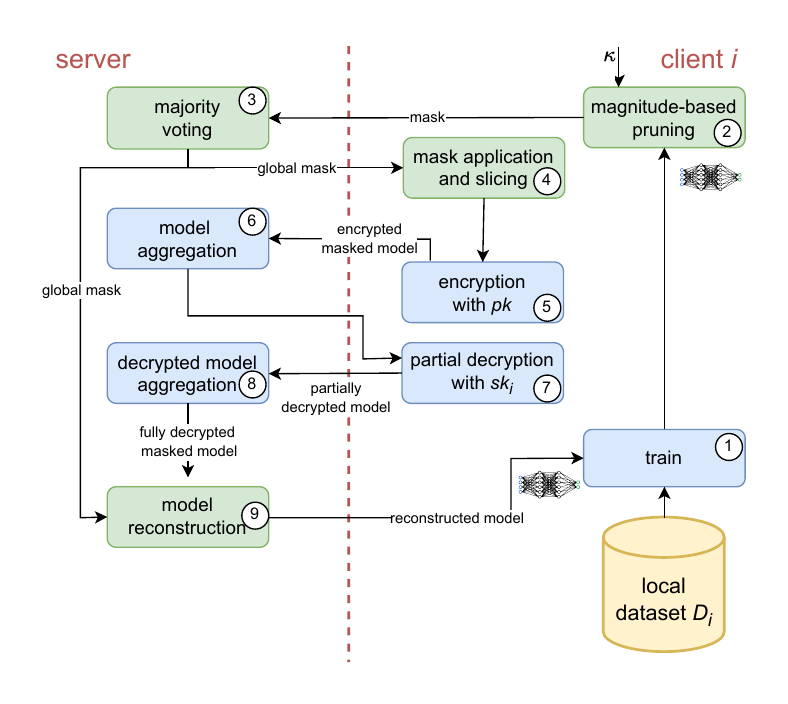}
    \caption{An illustration of MASER components and data flow between them}
    \label{fig:masertech}
\end{figure}

\subsection{Overview}\label{sec:overview}
MASER provides strong privacy guarantees by encrypting the reduced model updates that clients send to the aggregation server using MKHE. 
We show the workflow of MASER in Figure~\ref{fig:masertech} and describe these steps below:


\begin{enumerate}[wide, labelwidth=!, labelindent=0pt]
    \setcounter{enumi}{-1}
    \item \textbf{Key Generation and Model Initialization.} 
    Before the first round of FL,
    each participating client generates its key pair and performs public key aggregation as explained in Section~\ref{subsec:bg_mkhe}.
    The FL aggregation server randomly initializes the parameters of the machine learning model and sends them to all clients.
    
    \item \textbf{Model Training.} In the beginning of each training round, 
    every client receives the global model 
    from the aggregation server
    and performs model training on their local dataset $\mathcal{D}_i$ for $e$ local epochs, with $i$ being the index of this client.

    \item \textbf{Magnitude-based Pruning.} 
    Once a client completes local training, 
    it applies MP to the trained model to generate a local mask 
    for model sparsification. The local mask consists of binary values, with ``1"s indicating that the corresponding parameters are important and should be preserved and ``0"s indicating that they are not important and can be pruned.
    Specifically, the weights with a magnitude greater than a pruning threshold $\kappa$ and all biases are considered important.
    The local mask 
    is then sent to the server for aggregation.
    
    
    \item \textbf{Mask Aggregation through Majority Voting.} 
    Upon receiving the masks from all participating clients,
    the server employs a majority voting strategy to generate a global mask, 
    in which ``1"s correspond to parameters that are 
    deemed important (i.e., voted for) by at least $50\%$ of the clients.
    This consensus algorithm offers robustness to malicious clients.

    
    \item \textbf{Mask Application and Slicing.} 
    Each client receives the global mask 
    from the server and sparsifies their local model according to the global mask. The sparsification filters out the unimportant parameters in the local model.
    The remaining (important) parameters are then reshaped into lists, named \textit{slices}. All slices share the same length, determined by the security parameters.
    
    \item \textbf{Model Encryption.} Clients encrypt their slices using the aggregated public key $pk$. These ciphertexts are shared with the server to perform aggregation using homomorphic addition, preventing the server from seeing the plaintext version of the important model parameters. 
    
    \item \textbf{Encrypted Model Aggregation.} 
    The server receives the encrypted slices from all clients participating in model training, and performs model aggregation using homomorphic addition in the ciphertext space.
    
    \item \textbf{Partial Decryption with Client Secret Keys.} 
    The aggregated slices are then sent back to the clients for partial decryption, where each client $i$ uses their secret key $sk_i$ to partially decrypt the aggregated slices. The result is then sent to the server.
    
    \item \textbf{Model Decryption and Parameter Averaging.}
    The server receives the partially decrypted ciphertexts for the aggregated slices and combines them to fully decrypt the aggregated slices. The fully decrypted slices contain the summation of all important model weights and biases, which are subsequently divided by the total number of users participating in model training for averaging in plaintext. 
    This process ensures that neither the server nor any client gains full access to other clients' model updates, because a specific client's update cannot be inferred from the aggregated result. 
    
    \item \textbf{Model Reconstruction.} 
    The decrypted slices only contain the important weights and biases of the global model. 
    Therefore, before distributing the global model to the clients and starting the next training round, the server must convert the decrypted slices into the shape of a full model according to the global mask. The model reconstruction is achieved by placing the elements from the decrypted slices into their corresponding positions in the original model while setting the unimportant weights to zero. 
    The reconstructed global model is then distributed to the clients for the next round of model training.
    Although performing model reconstruction on the client side can reduce the communication cost, we perform it on the server side to minimize redundant computation and possibly allow a different set of clients to receive the full model and participate in the next round of model training. 
    In Section~\ref{sec:overhead}, we show that this approach does not significantly increase the communication cost as the overhead of transmitting the reconstructed model is negligible compared to the ciphertexts.

\end{enumerate}

\begin{algorithm}[!htb]
\small
\SetKwInOut{Execute}{Client $i$ Executes}
\KwData{Secure parameters $\mathcal{S}$, number of clients $m$, slice size $\lambda$, local training dataset $\mathcal{D}_i$, prunning threshold $\kappa$} 
\KwResult{Decrypted model parameters $\theta$}
\Execute{}
\Indp
    $sk_i$, $pk_i\leftarrow$ \texttt{KeyGen}($\mathcal{S}$) \tcp{Generate local key pair}
    Send $pk_i$ to key manager \tcp{To aggregate public keys}
    Receive aggregated public key $pk$: $pk_i \leftarrow pk$ \\ 
    \For{each round $t=1,2,…$} {
        Pull aggregated model parameters $\theta^t$ from the server: $\theta_i^t \leftarrow \theta^t$ \\ 
        \For {local epoch $e=1,2,...$} {
            $\theta_i^t \leftarrow$ \texttt{Train}($\theta_i^t$, $\mathcal{D}_{i}$) \tcp{Update local model}
        }
    $M_i^t \leftarrow$ \texttt{MaskGen}
    ($\theta_i^t$, $\kappa$) \tcp{Generate mask}
    Send $M_i^t$ to server \tcp{To aggregate masks}
    Receive aggregated mask and update local mask: $M_i^t \leftarrow M^t$ \\
    $v_i^t \leftarrow$ \texttt{Slice}($\theta_i^t \odot M_i^t, \lambda$) \tcp{Apply aggregated mask and prepare slices}
    $ct_i^t\leftarrow \texttt{Enc}(v_i^t, pk)$ \tcp{Encrypt slices using public key}
    Send $ct_i^t$ to server \\ 
    Receive $ct^t$ from server \tcp{$ct^t$: aggregated ciphertext}
    $pd_i^t\leftarrow$ \texttt{PartialDec}($ct^t$, $sk_i$) \tcp{Partially decrypt $ct^t$ using secret key}
    Send $pd_i^t$ to server 
    }
\Indm
\SetKwInOut{Execute}{Server Executes}
\Execute{}
\Indp
    Initialize $\theta^0$ \\
    \For{each round $t=1,2,…$} {
        Send model $\theta^t$ to client $i$ \\
        Receive local mask $M_i^t$ \\
        $M^t \leftarrow$ \texttt{MaskAgg}($M_1^t, \cdots, M_m^t$) \tcp{Perform mask aggregation}
        Send $M^t$ to client $i$ \\
        Receive $ct_i^t$ from clients \tcp{$ct_i^t$: ciphertext of slices}
        $ct^t \leftarrow \sum_{i=1}^m ct_i^t$ \tcp{Homomorphic parameter aggregation}
        Send $ct^t$ to client $i$ \\ 
        Receive $pd_i^t$ from clients \tcp{$pd_i^t$: partially decrypted ciphertext}
        $Q \leftarrow \texttt{Merge}(pd_1^t,\cdots,pd_m^t)$ \tcp{Aggregate partially decrypted ciphertext}
        $\theta^{t+1} \leftarrow \texttt{Reshape}(\texttt{Decode}(\frac{Q}{m}))$ \tcp{Decode and reshape $Q$ into plaintext model}
    }
\Indm
 \caption{MASER} 
 \label{alg:maser}
\end{algorithm}

 

\subsection{Model Training and Sparsification}\label{sec:sparsification}  

MASER uses MKCKKS-based MKHE~\cite{chen2019efficient} to safeguard the model parameters shared with the aggregation server in FL. 
In particular,
each client $i$ participating in MASER-based PPFL first generates their unique secret key $sk_i$ and public key $pk_i$. However, MKCKKS-based MKHE is not directly applicable in FL, because using these public keys for encryption may lead to privacy leakage during decryption~\cite{ma2022privacy}. 
To mitigate this, 
we adopt the key aggregation strategy proposed in xMKCKKS~\cite{ma2022privacy}. Specifically, the public keys are aggregated and encryption is done using this aggregate key by each individual client. 
Since $sk_i$ used for decryption is withheld by each client, 
other clients or the server cannot decrypt a client's model updates.

In this strategy, we employ a trusted key manager who aggregates the public keys shared by all clients to generate an aggregated public key $pk$. 
The $pk$ is then sent to all clients to replace their unique public key $pk_i$.
We show the pseudocode of the key generation process in Algorithm~\ref{alg:maser} from Line~1 to~3.
After the key generation process finishes, the aggregation server initializes the machine learning model $\theta^0$ and establishes a connection with each client to start the federated model training.\looseness=-1

\noindent \textbf{Training.}
MASER supports model training when the data held by the clients are IID and non-IID. 
In the beginning of a training round $t$, 
each client $i$ pulls the global model parameters, denoted as $\theta^t$, from the server to update their local model parameters, denoted as $\theta_i^t$. 
Then, they train their local model using their local dataset $\mathcal{D}_i$ for $e$ local epochs.
When the clients' local datasets are IID, MASER follows FedAvg~\cite{mcmahan2017communication} to train the local model $\theta_i^t$ by minimizing the loss function, i.e., cross-entropy loss, via the standard Adam optimizer.
However, FedAvg can perform poorly on non-IID data~\cite{zhao2018federated}. 
In this case, MASER supports local model training on non-IID data following  FedProx~\cite{li2020federated}, 
which adds a proximal term as introduced in~(\ref{eq:fedprox}) to the local loss function. This proximal term helps to manage heterogeneity in non-IID data by constraining the divergence of each client’s local model from the global model, thereby promoting better convergence across diverse datasets~\cite{li2020federated}.
Note that the aggregation of model parameters, via FedAvg or FedProx, is performed using homomorphic operations.
The pseudocode for model training is shown in Algorithm~\ref{alg:maser}, from Line~4 to~7.\looseness=-1

\vspace{.1cm}
\noindent \textit{Remark.} 
While our proof-of-concept employs FedAvg and FedProx for server-side aggregation, it is important to note that other aggregation rules that involve performing basic arithmetic operations on client updates (including robust and fair aggregation algorithms~\cite{karimireddy2020scaffold, wang2020tackling, hsu2019measuring}) can be easily adopted in MASER by
substituting these operations with the equivalent homomorphic operations.

\vspace{.1cm}
\noindent \textbf{Sparsification.}
MKHE-based FL approaches perform homomorphic encryption on the trained local model and then share the encrypted model updates with the server for aggregation. However, homomorphic encryption operations are computationally expensive, hence encrypting and transmitting the entire model in FL would make the overhead prohibitive.
MASER utilizes model sparsification to reduce the number of model parameters involved in the encryption and communication with the aggregation server.
Specifically, it uses MP for sparsification, pruning weights whose magnitudes are not among the top $\kappa$ percent of all weights in that model. 
The weights that are not pruned and all bias terms are considered \emph{important parameters}.
A local binary mask $M_i^t$ is then generated for the model parameterized by $\theta_i^t$ trained by client $i$, 
with ``1"s indicating important parameters and ``0"s corresponding to the pruned weights.
Since our pruning threshold is defined as a percentile, 
the smallest magnitude required for a parameter to be deemed important is different for each client and each round of pruning.

Instead of directly applying the local mask $M_i^t$ on each user's local model $\theta_i^t$, MASER requires clients to send their local binary masks to the server to generate a unified global mask $M^t$. 
This ensures that the model sparsification process is consistent across all clients, allowing the server to accurately aggregate the encrypted sparse models. Without a consistent global mask, the aggregated model would be misaligned, with different sparsity patterns across clients, leading to incorrect parameter aggregation. Note that the server receives the binary masks only,
and does not know the magnitudes of the important parameters.

For each training round, the server generates a global mask $M^t$ by using a majority vote-based aggregation.
In particular, the server performs element-wise summation for all local masks, then divides the result by the number of clients $m$, and thresholds it against 50\%. The mask aggregation rule is expressed as:
\begin{equation}
    M^t{(j,k)} =
\begin{cases}
    1, & \text{if } \frac{1}{m}\sum_{i=1}^m {M_i^t}{(j,k)} \geq 0.5, \\
    0, & \text{otherwise},
\end{cases}
\end{equation}
where $m$ is the total number of clients participating in model training, $t$ is the index for the current training round, $(j,k)$ is an index into the local mask $M_i^t$ or the global mask $M^t$.
A model parameter is deemed important, i.e., the corresponding element of the global mask is ``1" , 
when at least $50\%$ of clients had ``1" for this parameter in their local mask. 
The global mask $M^t$ is then shared with all clients to replace their local mask $M_i^t$.
Each client then performs element-wise multiplication of their updated local mask and the trained local model, i.e., $\widetilde{\theta}_i^t \leftarrow M_i^t \odot \theta_i^t$, where $\widetilde{\theta}_i^t$ is the sparsified local model for client $i$ at round $t$.
We note that the mask aggregation strategy also helps mitigate a mask poisoning attack by limiting how much a client's local mask can contribute to the global mask generation.
The pseudocode for mask generation and aggregation is shown in Algorithm~\ref{alg:maser}, between Line~8 and~10, and Line~20 and~22.

\subsection{Parameter Slicing for Efficient Homomorphic Encryption and Model Aggregation}\label{sec:slicing}
The amount of data that can be encrypted by the CKKS-based homomorphic encryption algorithm (i.e., the number of slots) is determined by the degree of the polynomial modulus $n$, where $n$ is a power of $2$. 
The number of slots is $\lambda = \frac{n}{2}$.
Therefore, to efficiently encrypt parameters of the sparsified model, we propose reshaping and dividing that model into multiple lists that share the same length of $\lambda$. We call these lists as \textit{slices}, and perform homomorphic encryption and model aggregation in the ciphertext space per slice.
In particular, we prepare the slices by starting with the model parameters in the first layer. The number of slices required by the first layer of model parameters is determined by $\lceil \frac{|\widetilde{\theta}_{(1)}|}{\lambda}\rceil$, where $|\widetilde{\theta}_{(1)}|$ is the number of important parameters in the first layer.
If the important parameters in the first layer are unable to fully fill the last slice, then the remaining slots in the last slice will be used to fill the important parameters in the second layer.
Hence a total of $\lceil \frac{|\widetilde{\theta}|}{\lambda}\rceil$ slices will be required to store all important model parameters.
Should there be any remaining slots available in the last slice after filling in all important model parameters, they will be set to zero.
This slicing technique allows efficient utilization of the limited slots and encrypting the sparsified model with a minimum amount of homomorphic encryption operations.
Once the slice preparation is complete, each client performs homomorphic encryption using their aggregated public key $pk$ for every slice and shares the encrypted ciphertext for every slice with the server to perform model aggregation in the ciphertext space. We denote the ciphertext encrypted by client $i$ at round $t$ as $ct_i^t$.
The pseudocode for slice preparation and homomorphic encryption is shown in Algorithm~\ref{alg:maser}, from Line~11 to~13.

Once the server receives the encrypted slices from all clients, it performs 
model aggregation by using homomorphic addition to sum up the ciphertext for each corresponding slice sent by all clients.
Then the summed ciphertexts $ct^t = \sum_{i=1}^m ct_i^t$ are sent back to all clients for partial decryption.
Note that in this step, we perform summation in the ciphertext space using homomorphic addition and defer the averaging to after the global model is fully decrypted.
The averaging step multiplies the summation of model parameters $ct^t$ by $\frac{1}{m}$, where $m$ is the number of clients participating in this round of model training.
We assume that the server has knowledge of $m$. Performing averaging in plaintext reduces the computational cost of homomorphic multiplication.
The pseudocode for model aggregation is given in Line~23 to~25 of Algorithm~\ref{alg:maser}.




\subsection{Decryption and Model Reconstruction}\label{sec:reconstruction}

The decryption of the ciphertext requires clients to perform partial decryption before the server can fully decrypt 
the aggregated ciphertext to plaintext~\cite{chen2019efficient}.
In the partial decryption phase, each client $i$ receives the aggregated ciphertext $ct^t$ from the server and uses their own secret key $sk_i$ to decrypt their own portion in the aggregated ciphertext. We use $pd_i^t$ to denote the ciphertext that is partially decrypted by client $i$ at round $t$.
$pd_i^t$ is then sent back to the server to fully decrypt the ciphertext. 
The steps for partial decryption are shown in Line~14 to~16 of Algorithm~\ref{alg:maser}.

The results after fully decryption are the slices summed over all clients in plaintext. We then divide each element in these slices by $m$ to obtain the aggregated model consisting of all important model parameters.
The server then reshapes the important parameters in the slices to obtain the original shape of the neural network by referring to the position of the corresponding important parameters in the global mask. This step is named model reconstruction.
Lastly, the fully decrypted and reshaped model parameters are sent to all clients to update their local models, and the next training round will begin. 
Although the server has access to the aggregated model, it is unable to access individual client's model update because the aggregation process combines all clients' updates into a single model update in a way that renders individual contributions indistinguishable.
Line~26 to~28 in Algorithm~\ref{alg:maser} outline the aggregation of the partially decrypted ciphertexts, full decryption of the ciphertexts, and model reconstruction. 

\section{Experimental Analysis}



\subsection{Datasets}
We evaluate MASER on two commonly used datasets, MNIST~\cite{lecun1998gradient} 
and CIFAR-10~\cite{krizhevsky2009learning}.
By testing MASER on these datasets, we demonstrate its effectiveness across different levels of task complexity.
We divide the training set of the two datasets among $m$ clients such that each client owns a non-overlapping portion of the training set with any other client. 
We consider both IID and non-IID settings when dividing the training set.
For the IID case, the training set of each dataset is split among all clients evenly.
For the non-IID case, we divide the training set according to the Dirichlet distribution and set all elements in the parameter vector $\alpha$ to $1.0$
to ensure variability across clients similar to~\cite{FRL}. The parameter $\alpha$ controls the degree of heterogeneity; lower values create more skewed class distributions, while higher values approach uniformity. Setting all elements in $\alpha$ to $1.0$ provides a moderate level of data heterogeneity, thereby ensuring each category preserves minimum representations.
This is similar to the non-IID setting studied in~\cite{fedphe, FRL}.
We evaluate the performance of the aggregated global model on the test set.

For performance evaluation, 
we use a modified version of LeNet-5 \cite{wortsman2020supermasks} and a Conv8 model \cite{ramanujan2020s} for the MNIST and CIFAR-10 datasets, respectively.
The modified LeNet-5 model for MNIST comprises 2 convolutional layers, followed by a pooling layer, and then 3 fully connected (FC) layers with mappings of 16$\times$4$\times$4 to 300, 300 to 100, and 100 to 10, respectively.
For CIFAR-10, we use a modified Conv8 architecture, referred to as ``ConvNet'' in this paper. This customized model consists of 7 convolutional layers followed by 3 fully connected layers, and no biases are applied in the network layers. The choice to remove biases is intended to simplify the model and reduce computational overhead in the federated setting.
We evaluate the aggregated model on the server side, using the test sets of MNIST and CIFAR-10 to measure the overall performance.

\subsection{Baselines}\label{sec:baseline}
\noindent \textbf{Vanilla FL.} 
We use FedAvg~\cite{mcmahan2017communication} in the IID case and FedProx~\cite{li2020federated} in the non-IID case for model aggregation without encryption. FedProx is a generalization of the FedAvg algorithm, which is meant to handle non-IID data across clients by adding a proximal term to stabilize training in heterogeneous settings. These two baseline methods establish the upper bound for model performance. 
However, they do not incorporate any privacy-preserving measures, leaving the shared model parameters unprotected.




\noindent \textbf{Federated Rank Learning (FRL)~\cite{FRL}.}
FRL is an approach in which clients send only the importance ranking of model weights in the neural network to the server (and not the weight values themselves). The server aggregates those rankings for model pruning.  FRL protects privacy in FL by eliminating the need to share model parameters. By comparing MASER with FRL, we aim to demonstrate the effectiveness of selectively encrypting important parameters in MASER in balancing privacy and model accuracy, as opposed to FRL's approach of rank-based aggregation.

\noindent \textbf{FedPHE~\cite{fedphe}.}
FedPHE employs a Single-key Homomorphic Encryption algorithm (SKHE) to encrypt the sparsified and packed model. A packed model means that multiple plaintext values from the model are combined into a single ciphertext, which reduces the number of ciphertexts to transmit and process. This packing approach helps minimize communication and computation overhead in FL by compressing the data before encryption. 
We use FedPHE as a baseline because a SKHE-based solution simplifies encryption, but requires all clients to share the same key. This poses privacy risks if any client is malicious or compromised. MASER, on the other hand, uses MKHE, allowing clients to have different keys, thereby enhancing privacy. Additionally, MASER selectively encrypts only the most important parameters to further reduce computational and communication costs compared to FedPHE, which encrypts a larger portion of the model. By comparing MASER with FedPHE, we highlight the efficiency gained through mask generation via majority voting and slicing,
while still maintaining stronger privacy guarantees.

\noindent\textbf{BatchCrypt~\cite{batchcrypt}.} 
BatchCrypt employs a multi-step approach to reduce the overhead of data transmission and encryption in FL. First, clients quantize their model parameters, which reduces precision but decreases the overall model size. Second, BatchCrypt applies packing, where multiple quantized values are combined into a single ciphertext, minimizing the number of ciphertexts needed for transmission. These packed values are then encrypted using a SKHE scheme before being sent to the server, which performs aggregation directly on the encrypted models. As BatchCrypt employs the SKHE scheme, it shares the same limitation as FedPHE, i.e., lack of robustness to malicious clients. BatchCrypt quantizes and encrypts the entire model, which can lead to greater precision loss and may not perform as efficiently in complex, non-IID data settings. In contrast, MASER selectively encrypts only the most important parameters. The comparison with this baseline helps understand the advantages of MASER in terms of privacy, efficiency, and accuracy.\looseness=-1

\noindent\textbf{MaskCrypt~\cite{maskcrypt}.}
MaskCrypt also utilizes SKHE and introduces a selective encryption technique for FL, aiming to balance privacy protection with efficiency by partially encrypting sensitive model parameters rather than the entire model. 
This partial encryption scheme uses static importance criteria for masking.
This opens an attack surface for malicious clients or adversaries to analyze the unencrypted model parameters and reconstruct client data via a data reconstruction attack~\cite{geiping2020inverting}. 
In contrast, the mask used in MASER for model sparsification is recalculated in every training round and the sparsified model parameters are fully encrypted. 
This provides stronger privacy guarantees against malicious clients and privacy attacks such as data reconstruction attacks. Comparing MASER with MaskCrypt highlights the added layer of security and adaptability provided by MASER's majority vote-based mask generation and its full encryption approach for critical parameters, which both mitigate privacy risks and maintain model performance in IID and non-IID settings.

\noindent\textbf{MKHE-based FL.}
This baseline is similar to MASER with one key difference: it does not apply the model sparsification technique to reduce the computation and communication overhead of MKHE.
Through comparison with this baseline, we show that the proposed sparsification method significantly reduces the overhead for MKHE-based FL. 

\noindent\textbf{Masking at Initialization (MAIN) with GraSP.}
This baseline replaces MASER's pruning technique with GraSP~\cite{wang2020picking}. Specifically, MAIN with GraSP uses the same MKHE algorithm and a majority voting-based mask consensus method as used in MASER, but employs GraSP~\cite{wang2020picking} for sparsification.
GraSP performs weight pruning only at initialization. Hence, the global mask generated in GraSP-based pruning is computed once at initialization and remains unchanged throughout the FL process. 
But the global mask is updated in MASER
in every training round.
Through comparison with this baseline,
we show that recalculating the local masks and updating the global mask in every round is necessary for enhanced performance.

\subsection{Implementation Details}

We implemented MASER on top of the Flower~\cite{beutel2020flower} Federated Learning framework, and built the machine learning models using PyTorch. MASER utilizes MKCKKS as the multi-key homomorphic encryption~\cite{kim2023asymptotically}. However, we only found a Golang-based implementation of MKCKKS\footnote{https://github.com/SNUCP/MKHE-KKLSS}. 
Since a Python-based MKCKKS library is necessary to integrate with the Python-based Flower framework and PyTorch, we implemented a Python wrapper for the Golang-based MKCKKS library. 
In particular, we designed a set of functions and data structures for the HE-related operations and used the \textit{cgo} library to export these functions into a C-style dynamic link library (DLL). 
Then we imported the DLL into Python and used the \textit{ctypes} library to convert the corresponding data structures and functions to Python. Note that \textit{cgo} does not support the map data structure used in the Golang implementation so we had to substitute the map data structure with Golang arrays.
Finally, we incorporated the public key aggregation method used in xMKCKKS~\cite{xmkckks-git} into our implementation.
We will publicly release the code for MASER upon acceptance of this paper. 

We simulated a cross-silo FL scenario with 
$m=5$
clients. 
For both datasets, we perform FL for $25$ global training rounds. In each global training round, clients train their local model for $e=5$ local epochs on their local datasets.
We set the modulus degree $\lambda$ for our MKHE algorithm to $8192$ and the learning rate of the machine learning models to $0.01$.
For non-IID data distribution, we set the parameter $\mu$ used in FedProx to $1.0$. 

We deployed MASER on two machines that are physically separate, located in two datacenters on the same campus, to simulate a real-world FL scenario incorporating network latency. We ran all clients on one machine and the server and key manager on the other machine.
The machine that hosts the clients is equipped with an Intel Core i9-9940X CPU, 128 GB of RAM, and an NVIDIA GeForce RTX 2080 Ti GPU to provide GPU acceleration for model training. 
The machine that hosts the server and key manager is equipped with an AMD EPYC 7313 CPU and 512 GB of RAM.


\subsection{Performance Evaluation}\label{sec:performance}
\begin{figure}[t!]
    \centering
    \includegraphics[clip, trim= 0cm 0cm 0cm 0.2cm, width=\linewidth]{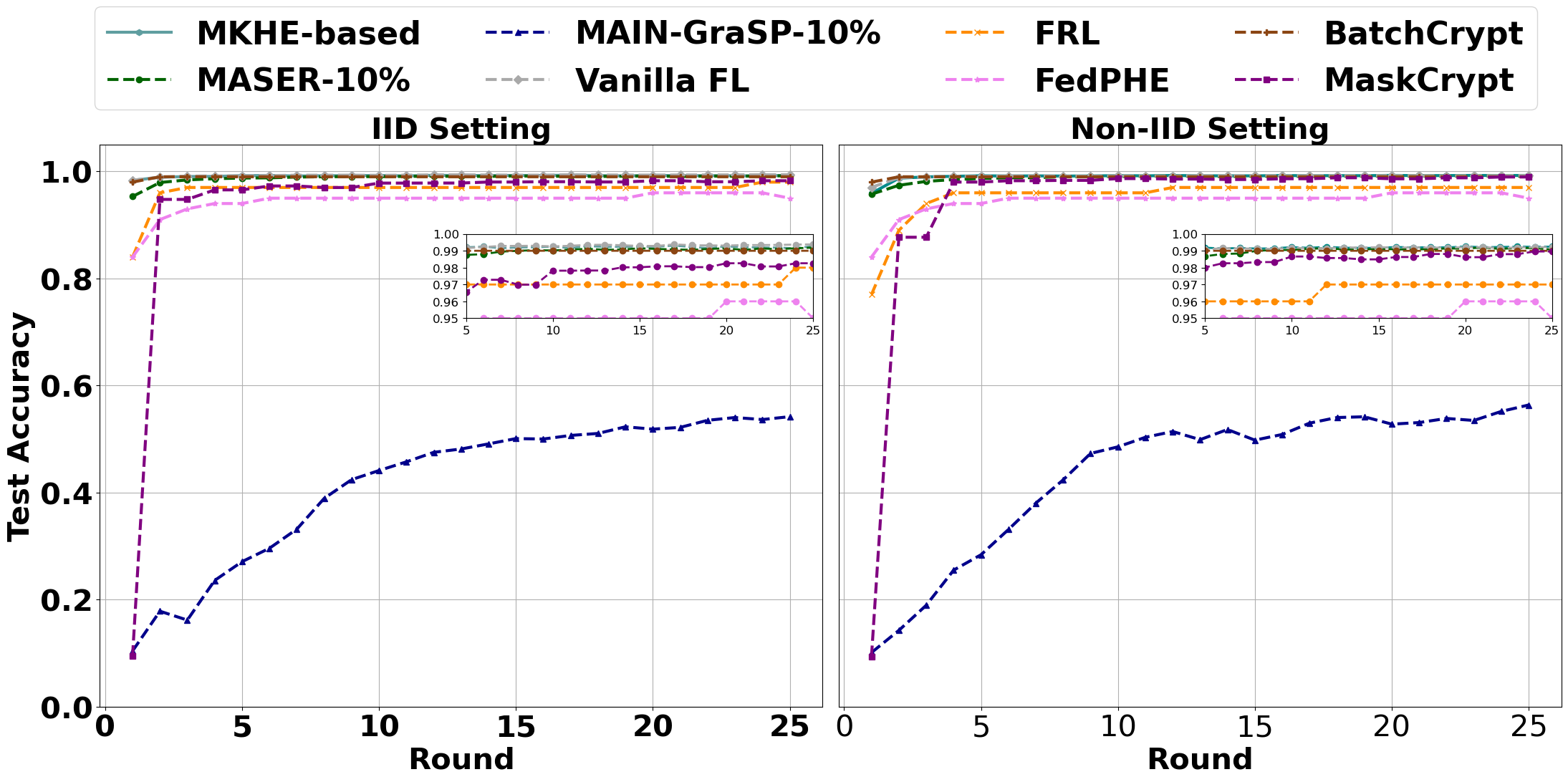}
    \caption{Test accuracy 
    across 25 FL rounds on MNIST}
    \label{fig:MNIST}
\end{figure}

\begin{figure}[t!]
    \centering
    \includegraphics[clip, trim= 0cm 0cm 0cm 0.2cm, width=\linewidth]{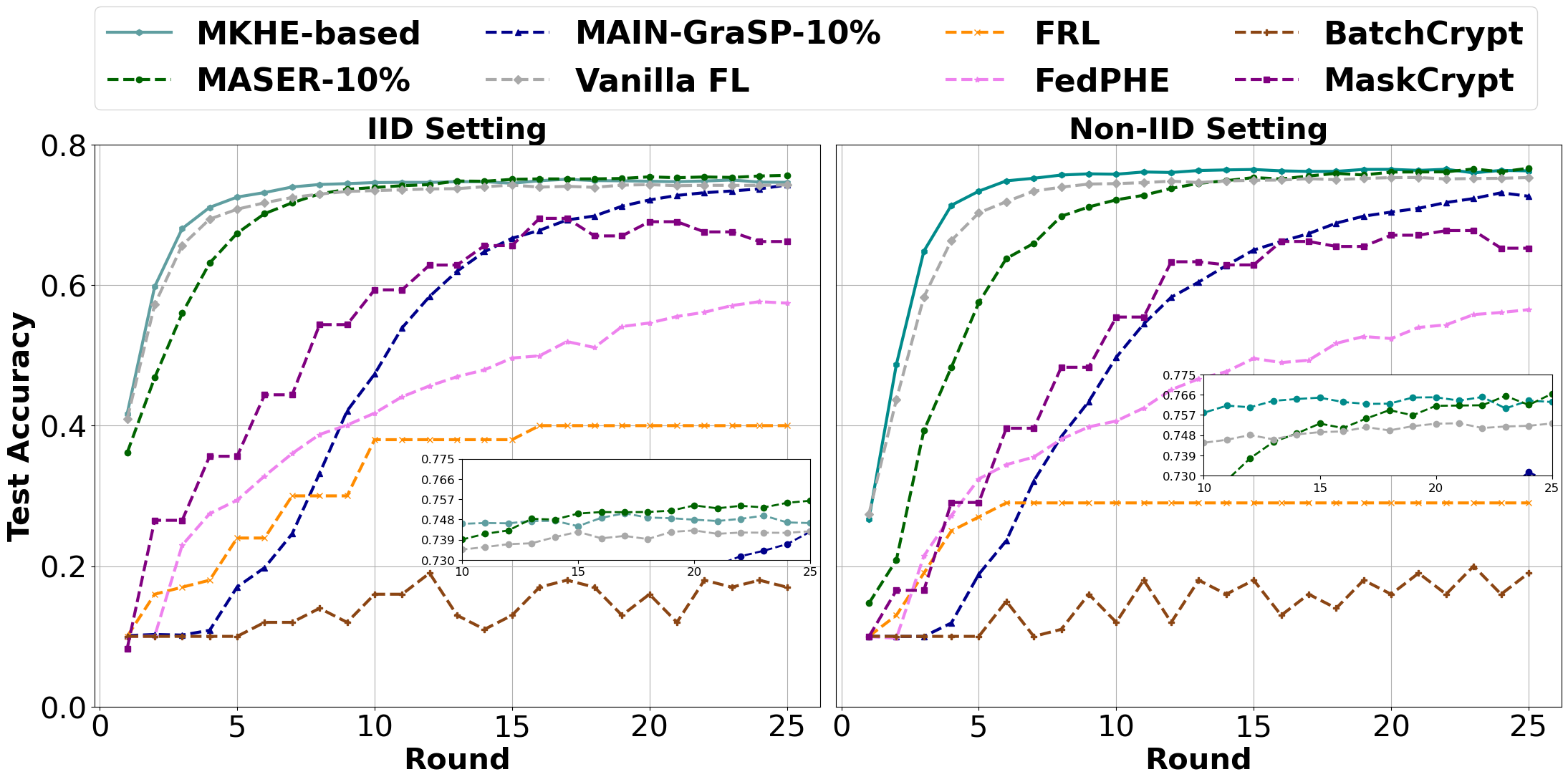}
    \caption{Test accuracy 
    across 25 FL rounds on CIFAR10}
    \label{fig:CIFAR10}
\end{figure}

We first evaluate the accuracy of the model trained by MASER on MNIST and CIFAR-10 datasets 
in both IID and non-IID settings.
Figure~\ref{fig:MNIST} and Figure~\ref{fig:CIFAR10} show the test accuracy (on the y-axis) measured across FL rounds (on the x-axis) for MASER and the baselines in IID and non-IID settings on MNIST and CIFAR-10 datasets, respectively. MASER and MAIN-GraSP choose the most important weights based on a predefined threshold. Here, MASER-10\% (MAIN-GraSP-10\%) refers to MASER (MAIN-GraSP) when we choose $\kappa = 10\%$ as the pruning threshold, meaning that only 10\% of the most important weights are not pruned (before majority voting). We have selected 10\% to demonstrate MASER's competitive performance while utilizing only 10\% of the parameters. Less aggressive pruning, which increases the overhead, only slightly increases the accuracy as shown in Appendix~\ref{app:A}. 


Figure~\ref{fig:MNIST} shows that all approaches except MAIN-GraSP-10\% converge to 95-99\% accuracy by round 5 in the IID setting and by round 10 in the non-IID setting on the MNIST dataset. 
In the IID setting, vanilla FL performs the best, reaching a stable accuracy of approximately 99.37\%, closely followed by MKHE-based FL and BatchCrypt, which achieve around 99-99.30\% accuracy. MASER-10\% achieves an accuracy of 99.18\%, only 0.19\% worse than vanilla FL. MaskCrypt also performs well, with an accuracy around 98.26\%, demonstrating robust performance although slightly lower than the leading methods.
In the non-IID setting, vanilla FL and MKHE-based FL perform the best reaching approximately 99.20\%. While MASER-10\% reaching 99.10\% closely followed by BatchCrypt and MaskCrypt which achieve around 98-99\% accuracy.
These results suggest that MASER-10\% shows comparable performance to the best baselines despite using only a subset of the model weights for aggregation in both IID and non-IID settings. The minor difference in accuracy can be attributed to MASER's pruning strategy, where less than 10\% (due to majority voting after keeping 10\% of parameters) of the most important model parameters are used for aggregation. The consistently high performance of MASER in the non-IID setting also demonstrates its ability to handle data heterogeneity.\looseness=-1



In contrast, MAIN-GraSP-10\% shows slower convergence, reaching an average accuracy of around 54.18\% by round 25 in the IID setting and 56.36\% by round 25 in the non-IID settings. This is because GraSP prunes a large portion of parameters at initialization, some of which would be among important parameters if pruning was delayed.  
This leads to reduced learning capacity and model performance. 
In contrast, MASER recalculates the pruning mask in each round, adapting to the evolving model. This adaptive approach in MASER allows for more effective parameter selection and retention, contributing to its superior performance compared to MAIN-GraSP. 

Figure~\ref{fig:CIFAR10} shows that MASER-10\% achieves 75.63\% accuracy by round 25 in the IID setting and 76.64\% by round 25 in the non-IID settings on the CIFAR dataset. These are the best results among all baselines at round 25. Vanilla FL and MKHE-based FL reach an accuracy level of approximately 74.27\% and 74.63\% in the IID setting, and 75.32\% and 76.27\% accuracy in the non-IID setting, respectively. These results highlight MASER’s effectiveness in parameter sparsification, slightly surpassing top-performing baseline methods while retaining only 10\% of the model weights. MAIN-GraSP-10\% reaches an accuracy level of approximately 74.25\% in the IID setting and 72.65\% in the non-IID setting by round 25. However, it converges much slower than MASER-10\%. Similar to the MNIST result, we believe that this is due to the extensive pruning done at initialization by GraSP.
These results confirm that MASER-10\% better accommodates changing data patterns, converges faster, and attains higher accuracy.

Among the other baseline methods, MaskCrypt reaches approximately 66.20\% accuracy in the IID setting and 65.30\% in the non-IID setting by round 25. FedPHE achieves around 57.44\% accuracy in the IID setting and 56.50\% accuracy in the non-IID setting by round 25. FRL and BatchCrypt perform significantly worse. FRL converges to around 40\% accuracy in the IID setting and to 29\% accuracy in the non-IID setting. BatchCrypt fluctuates below 20\% both in the IID and non-IID settings. These results demonstrate limitations in their ability to handle heterogeneous data distribution in the CIFAR-10 dataset when utilizing relatively simple models such as ConvNet. \looseness=-1

To summarize, MASER-10\% demonstrates consistent performance on par with the top-performing baselines across different data distributions, despite utilizing only 10\% of the most important weights.


\begin{figure}[t]
    \centering
    \includegraphics[clip, trim= 0cm 0cm 0cm 0.2cm, width=\linewidth]{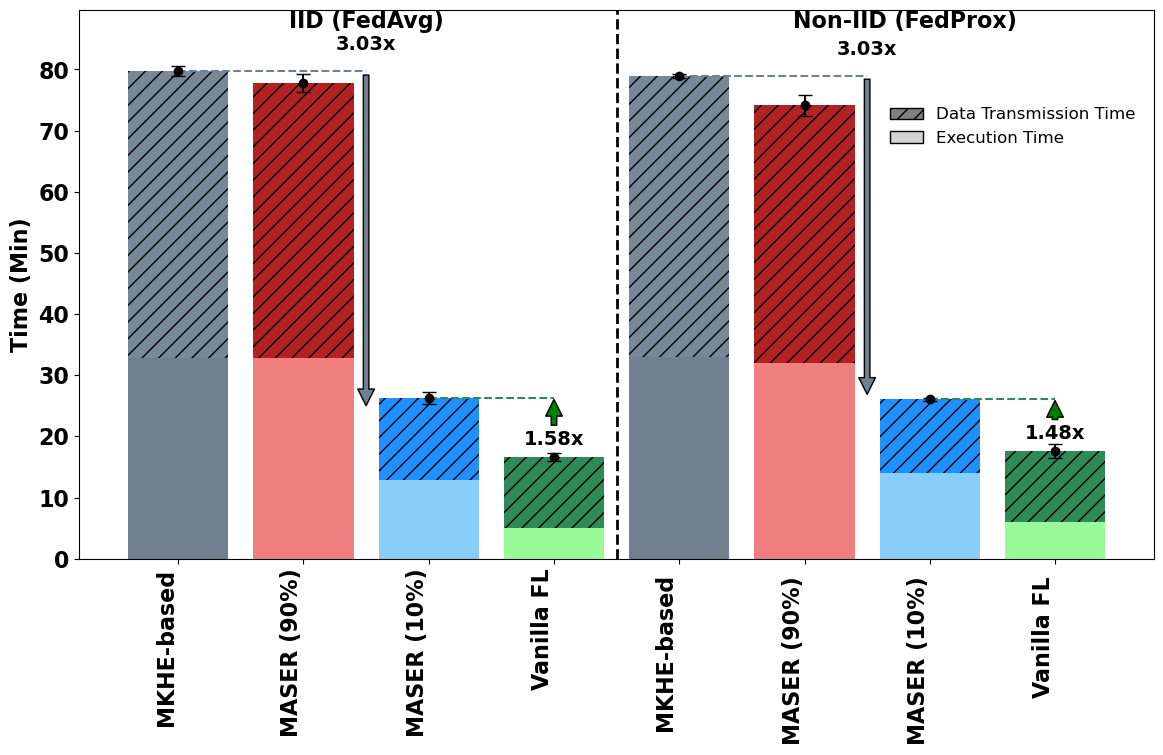}
    \caption{Time overhead for 25 FL rounds on MNIST}
    \label{fig:MNIST_time}
\end{figure}

\subsection{Overhead Analysis}\label{sec:overhead}

The main design goal of MASER is to minimize the overhead of MKHE-based FL while offering strong privacy guarantees. In this section, we restrict our analysis to the MKHE-based FL methods considering their computation and communication costs. 
The baseline methods that utilize a SKHE scheme are not directly comparable to MASER and the MKHE-based baseline
and are therefore disregarded here.
For example, many of these methods simulate the server and clients as processes that run on the same machine, while we deploy the server and clients on two separate machines, introducing additional communication latency. 
We analyze the overhead by running 25 FL rounds to evaluate the practicality of MASER for cross-silo FL. 
We run each experiment 5 times and report the average data transmission and execution time. The error bar shows the standard deviation, computed over the 5 runs.

\begin{figure}[t!]
    \centering
\includegraphics[clip, trim= 0cm 0cm 0cm 0.2cm, width=\linewidth]{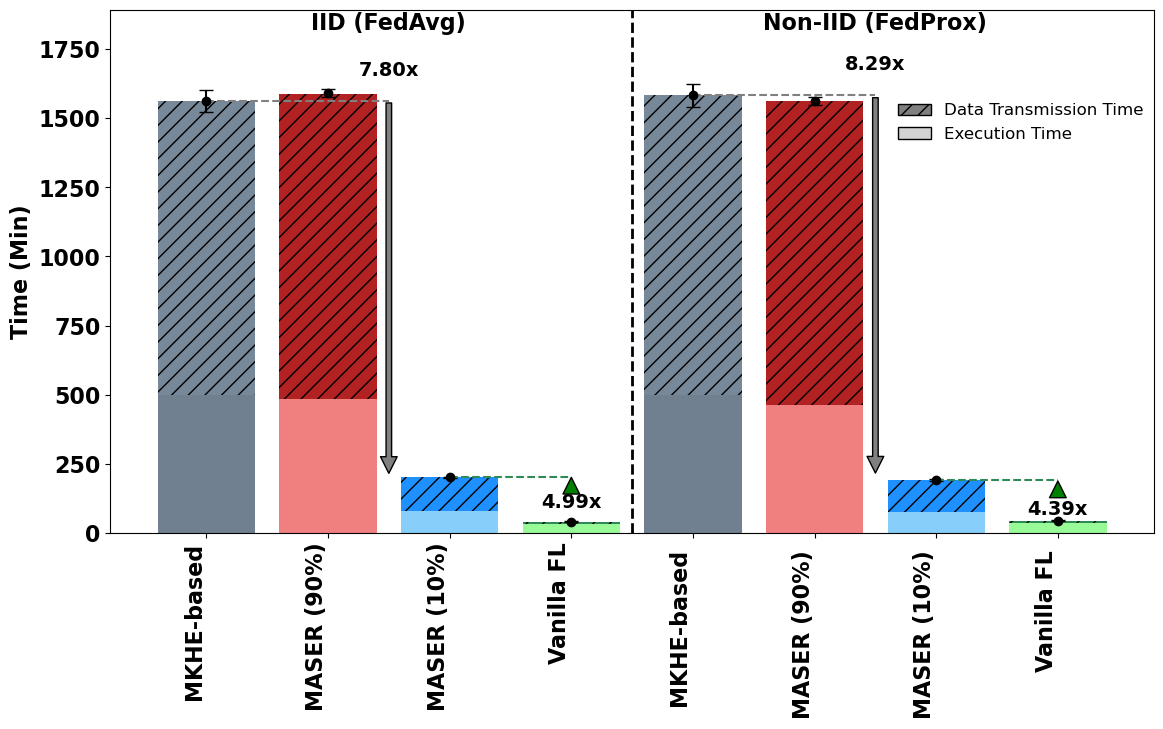}
    \caption{Time overhead for 25 FL rounds on CIFAR-10}
    \label{fig:CIFAR10_time}
\end{figure}

Figures \ref{fig:MNIST_time} and \ref{fig:CIFAR10_time} show the total overhead, measured in minutes, incurred by various federated learning strategies in IID and non-IID settings on MNIST and CIFAR-10, respectively. We break down the total running time into two parts:
data transmission time (hatched) and execution time (solid). We can readily see the efficiency of MASER-10\% relative to the MKHE-based FL baseline. 

\begin{table*}[!htb]
\centering
\caption{Size of the data after serialization (in MB)}
\begin{tabular}{|l|c|c|c|c|c|c|}
\hline
\textbf{Dataset} & \textbf{Threshold}  & \textbf{Per-client mask} & \textbf{Global mask} & \textbf{Encrypted slices} & \textbf{Aggregated slices (enc.)} & \textbf{Global model} \\ \hline
\textbf{CIFAR10 (ConvNet)} & 10\% & 11.67  & 23.34 & 62.85 & 188.56  & 23.33 \\ \hline
\textbf{CIFAR10 (ConvNet)} & 90\% & 11.67  & 23.34 & 633.92 & 1901.8 & 23.33 \\ \hline
\textbf{MNIST (modified LeNet-5)} & 10\% & 0.45  & 0.89 & 25.14 & 75.43 & 0.89 \\ \hline
\textbf{MNIST (modified LeNet-5)} & 90\% & 0.45  & 0.89 & 25.14 & 75.43  & 0.89 \\ \hline
\end{tabular}
\label{tab:serialization_size}
\end{table*}

We first look at the overhead analysis on the MNIST dataset in Figure~\ref{fig:MNIST_time}. As it can be seen, vanilla FL has the fastest running time, i.e., 16.67 minutes in the IID setting and 17.66 minutes in the non-IID setting. This is expected as it does not employ encryption and is vulnerable to privacy attacks. MKHE-based FL has the highest time overhead on average, i.e., 80 minutes in the IID setting, 58.79\% of which is for data transmission, 
and 78 minutes in the non-IID setting, 58.29\% of which is for data transmission. This shows that homomorphic operations are expensive computationally, and the encrypted ciphertext inflates the size of data to be transmitted, drastically increasing both the execution time and the data transmission time. \looseness=-1

By pruning only 10\% of the model weights, MASER-90\% slightly reduces the total run time by 1.95 minutes in the IID setting and by 4.75 minutes in the non-IID setting compared to MKHE-based FL. 
But MASER-10\% significantly reduces the time overhead (3.03${\times}$ reduction) 
compared to MKHE-based FL in both IID and non-IID settings. More precisely, it only requires 1.58${\times}$ the total running time of  FedAvg (vanilla FL) in the IID setting and 1.48${\times}$ the total running time of FedProx (vanilla FL) in the non-IID setting.

Figure~\ref{fig:CIFAR10_time} shows the overhead analysis on the CIFAR-10 dataset. 
MKHE-based FL and MASER-90\% yield the highest total run time, exceeding 1500 minutes in both settings. This significant overhead arises from the need to encrypt and transmit a large number of parameters in both cases. In contrast, MASER-10\% drastically reduces the total overhead
by 7.80${\times}$ in the IID setting and by 8.29${\times}$ in the non-IID setting, compared to MKHE-based FL. \looseness=-1
These results, along with analysis in Section~\ref{sec:performance}, reveal that MASER-10\% can maintain model performance and provide a strong privacy guarantee at the cost of a slight increase in the communication and computation overhead compared to vanilla FL. 
It is important to note that the total overhead of MASER-10\% is almost evenly divided by the data transmission time and execution time, suggesting that our model sparsification technique successfully minimizes both communication and computation overhead by reducing the total number of parameters to be encrypted and transmitted.

To understand how MASER minimizes the  communication and computation overhead,
we also report the size of data to be transmitted after serialization in megabytes (MB) at different FL stages for MASER-10\% and MASER-90\% in Table~\ref{tab:serialization_size}.
In particular, we consider the size of the serialized local mask, aggregated global mask, encrypted slices, aggregated slices,
and the decrypted global model parameters. 
Serialization, in this context, refers to the process of converting complex data structures, such as client masks, model parameters, and encrypted ciphertexts, into a format more suitable for transmission.
In the case of MNIST with the modified LeNet-5, the impact of the pruning threshold on the encrypted slice size is minimal due to the model's compact size. For both 10\% and 90\% pruning thresholds, all the parameters can fit into one slice, resulting in the same encrypted slice size of 25.14 MB and an aggregated ciphertext size of 75.43 MB. 
In contrast, CIFAR-10 with the ConvNet model shows a significant difference in the encrypted data size between the 10\% and 90\% pruning thresholds. At the 10\% threshold, only one slice is created, resulting in an encrypted slice size of 62.85 MB and an aggregated ciphertext size of 188.56 MB. However, at the 90\% threshold, the larger number of parameters requires 10 slices, raising the encrypted slice size to 633.92 MB and the aggregated ciphertext size to 1901.8 MB. This proportional increase demonstrates how the number of slices (and consequently the size of encrypted data) expands with higher thresholds.

\subsection{Privacy Leakage Analysis}
\begin{figure}[t]
    \centering
    \includegraphics[clip, trim= 2.5cm 4.5cm 2.5cm 2.5cm, width=.5\textwidth]{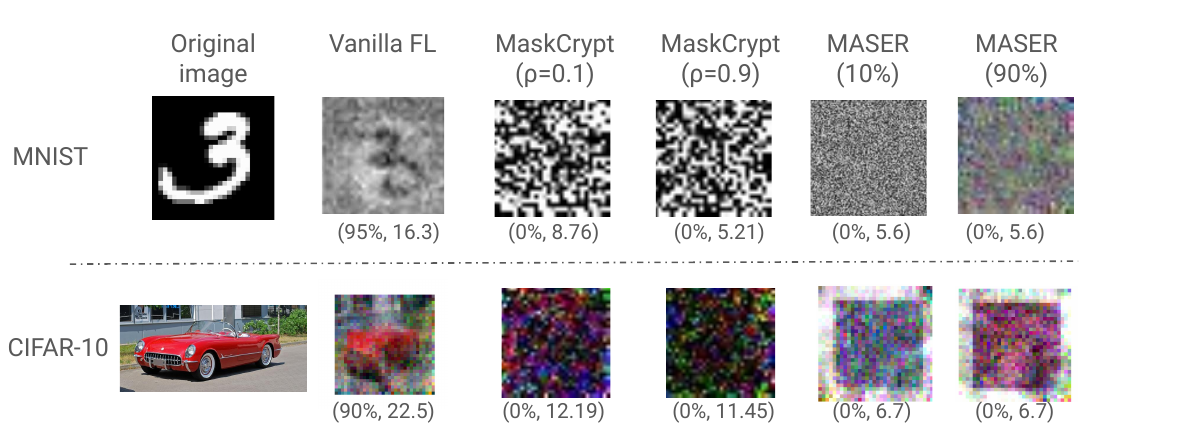}
    \caption{Original and reconstructed images via the data reconstruction attack on the FL model}
    \label{fig:GIA}
\end{figure}
We evaluate the privacy-preserving capability of MASER against a data reconstruction attack~\cite{zhu2019deep, geiping2020inverting, yin2021see, zhao2020idlg, wei2020framework} in which a client's private data is recovered from its model parameter update or gradient information. 
In particular, we follow the reconstruction attack described by Geiping \textit{et al.}~\cite{geiping2020inverting} to assess the potential privacy leakage from the shared model updates in FL. We measure the Attack Success Ratio (ASR) and Peak Signal-to-Noise Ratio (PSNR) of reconstructed images, with higher values indicating greater privacy leakage~\cite{geiping2020inverting}.
Figure~\ref{fig:GIA} shows results for MNIST and CIFAR-10 datasets across different FL methods, including vanilla FL, MaskCrypt with varying encryption levels $\rho$, and MASER with different pruning thresholds. The tuple below each reconstructed image indicates the ASR (\%) and PSNR value (in dB).
For MASER and MaskCrypt, the attack is performed on the fully and partially encrypted model update, respectively.\looseness=-1

As it is evident from the high success rate (95\% for MNIST and 90\% for CIFAR-10) and high PSNR value (16.3dB for MNIST and 22.5dB for CIFAR-10), there is significant privacy leakage in vanilla FL. In contrast, MASER, with 10\% and 90\% threshold, yields strong privacy protection, displaying near-zero ASR and low PSNR values (around 5.6dB for MNIST and 6.7dB for CIFAR-10), leaving no discernible pattern in the reconstructed image.
The MaskCrypt results are included for comparison. MaskCrypt employs selective encryption (based on the $\rho$ value) with a SKHE scheme. Although it does not exhibit significant privacy leakage at high encryption levels (e.g., $\rho=0.9$), it relies on SKHE, allowing a malicious client to decrypt the model updates of other clients
and achieve a high ASR as in vanilla FL.
Moreover, at low encryption levels (e.g., $\rho=0.04$ shown in Figure 6 of~\cite{maskcrypt}), MaskCrypt does not encrypt many model parameters, leaving them exposed to an adversary.
MASER, on the other hand, fully encrypts the parameters of the sparsified model, ensuring robustness to the reconstruction attack. The MKHE-based encryption also prevents malicious clients from circumventing the encryption of model updates.

\subsection{Robustness Analysis}
\begin{figure}[t!]
    \centering
    \includegraphics[clip, trim= 0cm 0cm 0cm 0.2cm, width=\linewidth]{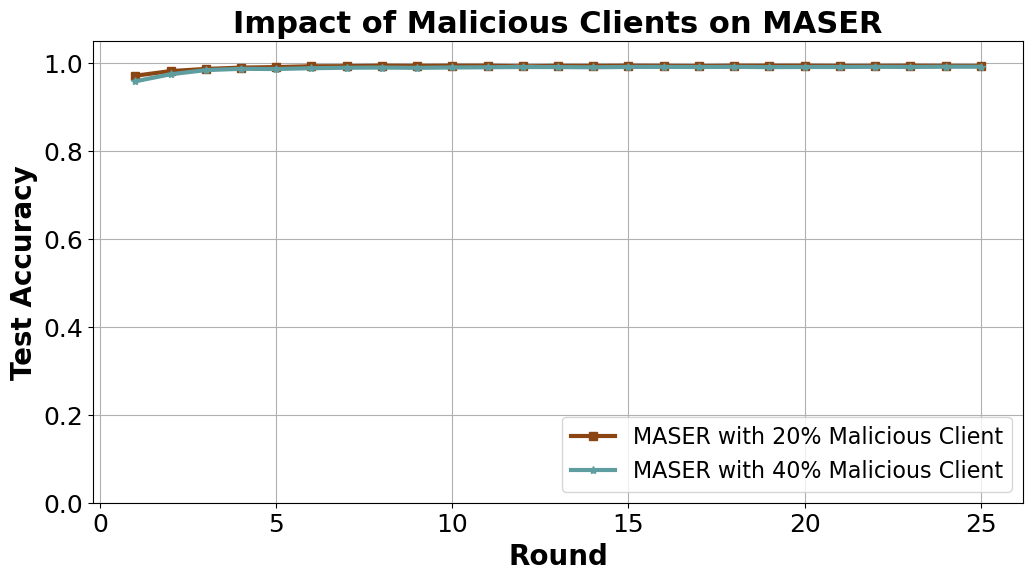}
    \caption{Test accuracy on MNIST with malicious clients}
    \label{fig:mal}
\end{figure}

Figure \ref{fig:mal} shows the robustness of MASER in the presence of malicious clients (not forming a majority), who attempt to compromise the training process by sending bogus masks to the server. 
We have considered two cases with 20\% and 40\% of clients acting maliciously.
It can be seen that MASER consistently maintains high and stable test accuracy throughout all rounds, showing negligible performance degradation even when 40\% of clients are malicious.

\section{Related Work}
Several privacy-preserving techniques have been integrated with FL. In this section, 
we mainly focus on the body of work that uses homomorphic encryption to protect data privacy in FL, as this is directly related to our work.

Data privacy is mainly protected in HE-based FL by allowing the server to perform operations (e.g., addition and multiplication) directly on encrypted data without decrypting it first~\cite{gentry2009fully, smart2010fully, al2023demystifying}. Single-key HE (SKHE) has been widely utilized in privacy-preserving FL~\cite{chen2019efficient, jin2023fedml, ma2022privacy, li2020homopai, xiong2024copifl, zhang2022homomorphic,maskcrypt}. However, a key limitation of SKHE in this context is that clients share a single set of encryption keys, which makes the entire system vulnerable as malicious clients could exploit the shared keys~\cite{10334062}.

Multi-key HE (MKHE) has recently gained popularity in FL due to its enhanced security, as each client has their own encryption keys. Cai et al. proposed a secure FL system using MKHE combined with a Trusted Execution Environment (TEE)~\cite{10334062}. They leverage a TEE to enhance security during computation on encrypted data while managing user dropout and privacy. However, they do not fully address the communication overhead introduced by MKHE. Similarly, Jing Ma et al. introduced xMKCKKS, a variant of MKHE that encrypts client updates to improve privacy~\cite{ma2022privacy}. Although their approach improves security compared to SKHE, it lacks mechanisms to reduce the overhead. 

Erfan et al. introduced a k-out-of-n secret sharing scheme allowing decryption without the need for all participants to be present~\cite{hosseini2021secure}. They also propose a compression technique using Random Linear Coding to reduce model size. However, the robustness of their scheme was not experimentally analyzed, and experiments were limited primarily to the MNIST dataset. Similarly, Park et al. proposed a distributed HE-based FL where each client uses different public keys~\cite{park2022privacy}. This approach enhances security by preventing key-sharing between clients, reducing the risk of a single compromised key to make the the entire system vulnerable. But, the authors did not include an analysis of the computational or communication overhead introduced by their approach. 
Instead, they primarily focused on the impact of varying key sizes on the system's running time, which provided limited insights into the trade-off between security and performance. While increasing the key size can improve security, it also significantly raises computation and data transmission costs. 
This is particularly important in FL, where maintaining an optimal balance between security and efficiency is crucial. In our work, we navigate this trade-off by ensuring that the key size remains large enough to guarantee security while still managing overhead, enabling scalable and efficient FL deployment in real-world applications.\looseness=-1

\textbf{Efficient HE.}  
One of the primary challenges in HE-based FL is the high communication and computation cost. 
To address this issue, several approaches have been proposed, focusing on optimizing HE by processing data in batches, applying masking techniques, or sparsifying the neural network~ \cite{batchcrypt, maskcrypt, fedphe, liu2023dhsa}. BatchCrypt \cite{batchcrypt} quantizes the model parameters and then packs them into smaller ciphertexts. Instead of encrypting each parameter individually, BatchCrypt processes parameters in groups (or batches), significantly reducing the amount of data sent between clients and the server. This batching technique lowers communication costs and speeds up the homomorphic encryption operations, allowing the server to efficiently perform operations on aggregated model updates. FedPHE employs a packing and encryption strategy based on the CKKS scheme, where sparsified model parameters are packed and encrypted to reduce both communication and computational overhead~\cite{fedphe}. The technique allows the system to tolerate stragglers (clients that are slower to respond or drop out) by encrypting only the most important parameters and skipping over less significant ones. By packing sparsified parameters into ciphertexts and only focusing on the essential data, FedPHE effectively balances security, efficiency, and robustness in federated learning. \cite{maskcrypt} introduced MaskCrypt, a masking technique to optimize the encryption process. MaskCrypt generates masks in such a way that when aggregated, the masks cancel each other out, allowing the server to retrieve the correct aggregated model without needing to handle all encrypted individual updates directly. MaskCrypt partially encrypts sensitive model parameters based on the generated masks. This method significantly reduces the computational burden on the server while preserving data privacy. The partial encryption of MaskCrypt may also leave certain model components exposed, leading to potential privacy vulnerabilities depending on the chosen thresholds~\cite{maskcrypt}. Importantly, BatchCrypt, FedPHE, and MaskCrypt rely on SKHE, which does not protect data privacy in the presence of malicious clients that can decrypt the updates of other clients and subsequently perform a data reconstruction attack.

Liu et al. proposed a Doubly Homomorphic Secure Aggregation (DHSA) scheme that employs a MKHE scheme, MKBFV, along with a Seed Homomorphic Pseudorandom Generator (SHPRG)~\cite{liu2023dhsa}. It improves both security and computational efficiency by generating masks for model updates instead of encrypting the entire model directly. 
Nevertheless, the use of SHPRG introduces potential risks from malicious clients. 
The security of SHPRG heavily relies on the integrity of the seed generation process. This opens up vulnerabilities, as malicious clients could attempt to influence the seed values, leading to compromised mask generation and, ultimately, privacy leakage~\cite{geiping2020inverting}. 
Our work addresses these concerns by not only focusing on enhancing the efficiency of MKHE but also ensuring robustness against privacy attacks by malicious clients.\looseness=-1

Although the above-mentioned methods reduce the communication and computational costs associated with HE-based FL, by increasing encryption efficiency, parameter pruning, and the selective transmission of important model updates, they have limitations such as susceptibility to privacy attacks and quantization errors.
Our approach addresses these issues by combining MKHE with a consensus-based pruning strategy and a slicing technique. 
Unlike the previous work that either trains a small model that has a limited learning capability,
or relies solely on packing strategies, we prune the model parameters before encryption, focusing only on the most critical parameters, and reconstruct the original model after aggregation and decryption. 
This reduces both computational and communication overhead,
and results in a balanced solution that maintains model performance, efficiency, and privacy in FL.

\textbf{Communication-Efficient FL.}
In FRL~\cite{FRL} clients do not transmit full model updates, instead they send their importance ranking of individual model parameters. The server aggregates these ranks to determine which subnetwork of the original network is a winning lottery ticket. This approach significantly reduces the communication overhead, as only the ranking is exchanged between clients and the server. 
Moreover, FRL improves privacy, as the model parameters are not directly transmitted and only importance scores are shared. This makes FRL highly effective for privacy-preserving and communication-efficient FL scenarios. 
However, as we showed in Section~\ref{sec:performance}, 
FRL needs many more rounds to achieve an acceptable level of accuracy in complex learning tasks.
This increases the total overhead of FRL, 
if we continue running it until it becomes competitive with other FL approaches. FedMask~\cite{fedmask} employs structured sparse binary masks to improve computational and communication efficiency of FL. While effective in reducing the overhead, such approaches assume a shared parameter space and overlook privacy concerns.
\section{Conclusion}

In this paper, we introduced MASER, a novel privacy-preserving, cross-silo federated learning approach that combines MKHE with model pruning, selective encryption, and slicing techniques to safeguard client data from an honest-but-curious server and malicious clients. Through an extensive experimental study and comparison with various baseline methods, we demonstrated that MASER significantly improves the performance and robustness of federated learning systems in both IID and non-IID settings. Our result shows that MASER not only achieves high accuracy (on par with vanilla federated learning algorithms) but also substantially reduces the computation and communication overhead introduced by privacy-preserving techniques. It also suggests that MASER is well-suited for deployment in real-world scenarios, where privacy and efficiency are key concerns. We believe that this work offers a promising pathway for developing more efficient, secure, and scalable federated learning systems.
\bibliographystyle{IEEEtran}
\bibliography{sample-base}

\begin{thebibliography}{10}
\providecommand{\url}[1]{#1}
\csname url@samestyle\endcsname
\providecommand{\newblock}{\relax}
\providecommand{\bibinfo}[2]{#2}
\providecommand{\BIBentrySTDinterwordspacing}{\spaceskip=0pt\relax}
\providecommand{\BIBentryALTinterwordstretchfactor}{4}
\providecommand{\BIBentryALTinterwordspacing}{\spaceskip=\fontdimen2\font plus
\BIBentryALTinterwordstretchfactor\fontdimen3\font minus
  \fontdimen4\font\relax}
\providecommand{\BIBforeignlanguage}[2]{{%
\expandafter\ifx\csname l@#1\endcsname\relax
\typeout{** WARNING: IEEEtran.bst: No hyphenation pattern has been}%
\typeout{** loaded for the language `#1'. Using the pattern for}%
\typeout{** the default language instead.}%
\else
\language=\csname l@#1\endcsname
\fi
#2}}
\providecommand{\BIBdecl}{\relax}
\BIBdecl

\bibitem{kairouz2021advances}
P.~Kairouz \emph{et~al.}, ``Advances and open problems in federated learning,''
  \emph{Foundations and trends in machine learning}, vol.~14, no. 1--2, pp.
  1--210, 2021.

\bibitem{zhu2019deep}
L.~Zhu, Z.~Liu, and S.~Han, ``Deep leakage from gradients,'' \emph{Advances in
  neural information processing systems}, vol.~32, 2019.

\bibitem{yin2021see}
H.~Yin, A.~Mallya, A.~Vahdat, J.~M. Alvarez, J.~Kautz, and P.~Molchanov, ``See
  through gradients: Image batch recovery via gradinversion,'' in
  \emph{Proceedings of the IEEE/CVF Conference on Computer Vision and Pattern
  Recognition}, 2021, pp. 16\,337--16\,346.

\bibitem{geiping2020inverting}
J.~Geiping, H.~Bauermeister, H.~Dr{\"o}ge, and M.~Moeller, ``Inverting
  gradients-how easy is it to break privacy in federated learning?''
  \emph{Advances in neural information processing systems}, vol.~33, pp.
  16\,937--16\,947, 2020.

\bibitem{wei2020federated}
K.~Wei \emph{et~al.}, ``Federated learning with differential privacy:
  Algorithms and performance analysis,'' \emph{IEEE Transactions on Information
  Forensics and Security}, vol.~15, pp. 3454--3469, 2020.

\bibitem{bonawitz2017practical}
K.~Bonawitz, V.~Ivanov, B.~Kreuter, A.~Marcedone, H.~B. McMahan, S.~Patel,
  D.~Ramage, A.~Segal, and K.~Seth, ``Practical secure aggregation for
  privacy-preserving machine learning,'' in \emph{proceedings of the 2017 ACM
  SIGSAC Conference on Computer and Communications Security}, 2017, pp.
  1175--1191.

\bibitem{batchcrypt}
C.~Zhang, S.~Li, J.~Xia, W.~Wang, F.~Yan, and Y.~Liu, ``Batchcrypt: Efficient
  homomorphic encryption for cross-silo federated learning,'' in
  \emph{Proceedings of the 2020 USENIX Annual Technical Conference (USENIX ATC
  2020)}, 2020.

\bibitem{jin2023fedml}
W.~Jin, Y.~Yao, S.~Han, C.~Joe-Wong, S.~Ravi, S.~Avestimehr, and C.~He,
  ``Fed{ML}-{HE}: An efficient homomorphic-encryption-based privacy-preserving
  federated learning system,'' in \emph{International Workshop on Federated
  Learning in the Age of Foundation Models in Conjunction with NeurIPS 2023},
  2023.

\bibitem{tramer2021differentially}
F.~Tramer and D.~Boneh, ``Differentially private learning needs better features
  (or much more data),'' in \emph{International Conference on Learning
  Representations}, 2021.

\bibitem{yang2019federated}
Q.~Yang, Y.~Liu, T.~Chen, and Y.~Tong, ``Federated machine learning: Concept
  and applications,'' \emph{ACM Transactions on Intelligent Systems and
  Technology (TIST)}, vol.~10, no.~2, pp. 1--19, 2019.

\bibitem{rivest1978data}
R.~L. Rivest, L.~Adleman, M.~L. Dertouzos \emph{et~al.}, ``On data banks and
  privacy homomorphisms,'' \emph{Foundations of secure computation}, vol.~4,
  no.~11, pp. 169--180, 1978.

\bibitem{fedphe}
N.~Yan, Y.~Li, J.~Chen, X.~Wang, J.~Hong, K.~He, and W.~Wang, ``Efficient and
  straggler-resistant homomorphic encryption for heterogeneous federated
  learning,'' in \emph{Proc. IEEE INFOCOM}, 2024.

\bibitem{maskcrypt}
C.~Hu and B.~Li, ``Maskcrypt: Federated learning with selective homomorphic
  encryption,'' \emph{IEEE Transactions on Dependable and Secure Computing},
  2024.

\bibitem{10334062}
Y.~Cai, W.~Ding, Y.~Xiao, Z.~Yan, X.~Liu, and Z.~Wan, ``Secfed: A secure and
  efficient federated learning based on multi-key homomorphic encryption,''
  \emph{IEEE Transactions on Dependable and Secure Computing}, vol.~21, no.~04,
  pp. 3817--3833, jul 2024.

\bibitem{ma2022privacy}
J.~Ma, S.-A. Naas, S.~Sigg, and X.~Lyu, ``Privacy-preserving federated learning
  based on multi-key homomorphic encryption,'' \emph{International Journal of
  Intelligent Systems}, vol.~37, no.~9, pp. 5880--5901, 2022.

\bibitem{park2022privacy}
J.~Park, N.~Y. Yu, and H.~Lim, ``Privacy-preserving federated learning using
  homomorphic encryption with different encryption keys,'' in \emph{2022 13th
  International Conference on Information and Communication Technology
  Convergence (ICTC)}.\hskip 1em plus 0.5em minus 0.4em\relax IEEE, 2022, pp.
  1869--1871.

\bibitem{li2020federated}
T.~Li, A.~K. Sahu, M.~Zaheer, M.~Sanjabi, A.~Talwalkar, and V.~Smith,
  ``Federated optimization in heterogeneous networks,'' \emph{Proceedings of
  Machine learning and systems}, vol.~2, pp. 429--450, 2020.

\bibitem{Han2015}
S.~Han, J.~Pool, J.~Tran, and W.~Dally, ``Learning both weights and connections
  for efficient neural network,'' in \emph{Advances in Neural Information
  Processing Systems}, vol.~28.\hskip 1em plus 0.5em minus 0.4em\relax Curran
  Associates, Inc., 2015.

\bibitem{FRL}
H.~Mozaffari, V.~Shejwalkar, and A.~Houmansadr, ``Every vote counts:
  {Ranking-Based} training of federated learning to resist poisoning attacks,''
  in \emph{32nd USENIX Security Symposium (USENIX Security 23)}.\hskip 1em plus
  0.5em minus 0.4em\relax USENIX Association, Aug. 2023, pp. 1721--1738.

\bibitem{huang2022cross}
C.~Huang, J.~Huang, and X.~Liu, ``Cross-silo federated learning: Challenges and
  opportunities,'' \emph{arXiv preprint arXiv:2206.12949}, 2022.

\bibitem{amari1993backpropagation}
S.-i. Amari, ``Backpropagation and stochastic gradient descent method,''
  \emph{Neurocomputing}, vol.~5, no. 4-5, pp. 185--196, 1993.

\bibitem{mcmahan2017communication}
B.~McMahan, E.~Moore, D.~Ramage, S.~Hampson, and B.~A. y~Arcas,
  ``Communication-efficient learning of deep networks from decentralized
  data,'' in \emph{Artificial intelligence and statistics}.\hskip 1em plus
  0.5em minus 0.4em\relax PMLR, 2017, pp. 1273--1282.

\bibitem{zhao2018federated}
Y.~Zhao, M.~Li, L.~Lai, N.~Suda, D.~Civin, and V.~Chandra, ``Federated learning
  with non-iid data,'' \emph{arXiv preprint arXiv:1806.00582}, 2018.

\bibitem{lyubashevsky2013ideal}
V.~Lyubashevsky, C.~Peikert, and O.~Regev, ``On ideal lattices and learning
  with errors over rings,'' \emph{Journal of the ACM (JACM)}, vol.~60, no.~6,
  pp. 1--35, 2013.

\bibitem{chen2019efficient}
H.~Chen, W.~Dai, M.~Kim, and Y.~Song, ``Efficient multi-key homomorphic
  encryption with packed ciphertexts with application to oblivious neural
  network inference,'' in \emph{Proceedings of the 2019 ACM SIGSAC Conference
  on Computer and Communications Security}, 2019, pp. 395--412.

\bibitem{Hassibi92}
B.~Hassibi and D.~Stork, ``Second order derivatives for network pruning:
  Optimal brain surgeon,'' in \emph{Advances in Neural Information Processing
  Systems}, vol.~5.\hskip 1em plus 0.5em minus 0.4em\relax Morgan-Kaufmann,
  1992.

\bibitem{LeCun89}
Y.~LeCun, J.~Denker, and S.~Solla, ``Optimal brain damage,'' in \emph{Advances
  in Neural Information Processing Systems}, vol.~2.\hskip 1em plus 0.5em minus
  0.4em\relax Morgan-Kaufmann, 1989.

\bibitem{park2020lookahead}
\BIBentryALTinterwordspacing
S.~Park, J.~Lee, S.~Mo, and J.~Shin, ``Lookahead: A far-sighted alternative of
  magnitude-based pruning,'' in \emph{International Conference on Learning
  Representations}, 2020. [Online]. Available:
  \url{https://openreview.net/forum?id=ryl3ygHYDB}
\BIBentrySTDinterwordspacing

\bibitem{Gale19}
\BIBentryALTinterwordspacing
T.~Gale, E.~Elsen, and S.~Hooker, ``The state of sparsity in deep neural
  networks,'' \emph{arXiv e-prints}, vol. arXiv:1902.09574, 2019. [Online].
  Available: \url{https://arxiv.org/abs/1902.09574}
\BIBentrySTDinterwordspacing

\bibitem{lee2018snip}
N.~Lee, T.~Ajanthan, and P.~H. Torr, ``Snip: Single-shot network pruning based
  on connection sensitivity,'' \emph{arXiv preprint arXiv:1810.02340}, 2018.

\bibitem{wang2020picking}
C.~Wang, G.~Zhang, and R.~Grosse, ``Picking winning tickets before training by
  preserving gradient flow,'' \emph{arXiv preprint arXiv:2002.07376}, 2020.

\bibitem{tanaka2020pruning}
H.~Tanaka, D.~Kunin, D.~L. Yamins, and S.~Ganguli, ``Pruning neural networks
  without any data by iteratively conserving synaptic flow,'' \emph{Advances in
  neural information processing systems}, vol.~33, pp. 6377--6389, 2020.

\bibitem{cheng2024surveypruning}
H.~Cheng, M.~Zhang, and J.~Q. Shi, ``A survey on deep neural network pruning:
  Taxonomy, comparison, analysis, and recommendations,'' \emph{IEEE
  Transactions on Pattern Analysis and Machine Intelligence}, 2024.

\bibitem{babakniya2023revisiting}
S.~Babakniya, S.~Kundu, S.~Prakash, Y.~Niu, and S.~Avestimehr, ``Revisiting
  sparsity hunting in federated learning: Why does sparsity consensus matter?''
  \emph{Transactions on Machine Learning Research}, 2023.

\bibitem{gez2023masked}
T.~L. Gez and K.~Cohen, ``A masked pruning approach for dimensionality
  reduction in communication-efficient federated learning systems,''
  \emph{arXiv preprint arXiv:2312.03889}, 2023.

\bibitem{aono2017privacy}
Y.~Aono, T.~Hayashi, L.~Wang, S.~Moriai \emph{et~al.}, ``Privacy-preserving
  deep learning via additively homomorphic encryption,'' \emph{IEEE
  Transactions on Information Forensics and Security}, vol.~13, no.~5, pp.
  1333--1345, 2017.

\bibitem{3133982}
K.~Bonawitz \emph{et~al.}, ``Practical secure aggregation for
  privacy-preserving machine learning,'' in \emph{Proceedings of the 2017 ACM
  SIGSAC Conference on Computer and Communications Security}, ser. CCS
  '17.\hskip 1em plus 0.5em minus 0.4em\relax ACM, 2017, pp. 1175--1191.

\bibitem{karimireddy2020scaffold}
S.~P. Karimireddy, S.~Kale, M.~Mohri, S.~Reddi, S.~Stich, and A.~T. Suresh,
  ``Scaffold: Stochastic controlled averaging for federated learning,'' in
  \emph{International conference on machine learning}.\hskip 1em plus 0.5em
  minus 0.4em\relax PMLR, 2020, pp. 5132--5143.

\bibitem{wang2020tackling}
J.~Wang, Q.~Liu, H.~Liang, G.~Joshi, and H.~V. Poor, ``Tackling the objective
  inconsistency problem in heterogeneous federated optimization,''
  \emph{Advances in neural information processing systems}, vol.~33, pp.
  7611--7623, 2020.

\bibitem{hsu2019measuring}
T.-M.~H. Hsu, H.~Qi, and M.~Brown, ``Measuring the effects of non-identical
  data distribution for federated visual classification,'' \emph{arXiv preprint
  arXiv:1909.06335}, 2019.

\bibitem{lecun1998gradient}
Y.~LeCun, L.~Bottou, Y.~Bengio, and P.~Haffner, ``Gradient-based learning
  applied to document recognition,'' \emph{Proceedings of the IEEE}, vol.~86,
  no.~11, pp. 2278--2324, 1998.

\bibitem{krizhevsky2009learning}
A.~Krizhevsky, G.~Hinton \emph{et~al.}, ``Learning multiple layers of features
  from tiny images,'' 2009.

\bibitem{wortsman2020supermasks}
M.~Wortsman, V.~Ramanujan, R.~Liu, A.~Kembhavi, M.~Rastegari, J.~Yosinski, and
  A.~Farhadi, ``Supermasks in superposition,'' \emph{Advances in Neural
  Information Processing Systems}, vol.~33, pp. 15\,173--15\,184, 2020.

\bibitem{ramanujan2020s}
V.~Ramanujan, M.~Wortsman, A.~Kembhavi, A.~Farhadi, and M.~Rastegari, ``What's
  hidden in a randomly weighted neural network?'' in \emph{Proceedings of the
  IEEE/CVF conference on computer vision and pattern recognition}, 2020, pp.
  11\,893--11\,902.

\bibitem{beutel2020flower}
D.~J. Beutel, T.~Topal, A.~Mathur, X.~Qiu, J.~Fernandez-Marques, Y.~Gao,
  L.~Sani, K.~H. Li, T.~Parcollet, P.~P.~B. de~Gusm{\~a}o \emph{et~al.},
  ``Flower: A friendly federated learning research framework,'' \emph{arXiv
  preprint arXiv:2007.14390}, 2020.

\bibitem{kim2023asymptotically}
T.~Kim, H.~Kwak, D.~Lee, J.~Seo, and Y.~Song, ``Asymptotically faster multi-key
  homomorphic encryption from homomorphic gadget decomposition,'' in
  \emph{Proceedings of the 2023 ACM SIGSAC Conference on Computer and
  Communications Security}, 2023, pp. 726--740.

\bibitem{xmkckks-git}
``xmkckks implementation,''
  \url{https://github.com/MetisPrometheus/MSc-thesis-xmkckks}, 2024 [Online],
  accessed in 2024.

\bibitem{zhao2020idlg}
B.~Zhao, K.~R. Mopuri, and H.~Bilen, ``{iDLG}: Improved deep leakage from
  gradients,'' \emph{arXiv preprint arXiv:2001.02610}, 2020.

\bibitem{wei2020framework}
W.~Wei, L.~Liu, M.~Loper, K.-H. Chow, M.~E. Gursoy, S.~Truex, and Y.~Wu, ``A
  framework for evaluating gradient leakage attacks in federated learning,''
  \emph{arXiv preprint arXiv:2004.10397}, 2020.

\bibitem{gentry2009fully}
C.~Gentry, ``Fully homomorphic encryption using ideal lattices,'' in
  \emph{Proceedings of the forty-first annual ACM symposium on Theory of
  computing}, 2009, pp. 169--178.

\bibitem{smart2010fully}
N.~P. Smart and F.~Vercauteren, ``Fully homomorphic encryption with relatively
  small key and ciphertext sizes,'' in \emph{Public Key Cryptography--PKC 2010:
  13th International Conference on Practice and Theory in Public Key
  Cryptography, Paris, France, May 26-28, 2010. Proceedings 13}.\hskip 1em plus
  0.5em minus 0.4em\relax Springer, 2010, pp. 420--443.

\bibitem{al2023demystifying}
A.~Al~Badawi and Y.~Polyakov, ``Demystifying bootstrapping in fully homomorphic
  encryption,'' \emph{Cryptology ePrint Archive}, 2023.

\bibitem{li2020homopai}
Q.~Li, Z.~Huang, W.-j. Lu, C.~Hong, H.~Qu, H.~He, and W.~Zhang, ``Homopai: A
  secure collaborative machine learning platform based on homomorphic
  encryption,'' in \emph{2020 IEEE 36th International Conference on Data
  Engineering (ICDE)}.\hskip 1em plus 0.5em minus 0.4em\relax IEEE, 2020, pp.
  1713--1717.

\bibitem{xiong2024copifl}
R.~Xiong, W.~Ren, S.~Zhao, J.~He, Y.~Ren, K.-K.~R. Choo, and G.~Min, ``Copifl:
  A collusion-resistant and privacy-preserving federated learning crowdsourcing
  scheme using blockchain and homomorphic encryption,'' \emph{Future Generation
  Computer Systems}, vol. 156, pp. 95--104, 2024.

\bibitem{zhang2022homomorphic}
L.~Zhang, J.~Xu, P.~Vijayakumar, P.~K. Sharma, and U.~Ghosh, ``Homomorphic
  encryption-based privacy-preserving federated learning in iot-enabled
  healthcare system,'' \emph{IEEE Transactions on Network Science and
  Engineering}, vol.~10, no.~5, pp. 2864--2880, 2022.

\bibitem{hosseini2021secure}
E.~Hosseini and A.~Khisti, ``Secure aggregation in federated learning via
  multiparty homomorphic encryption,'' in \emph{2021 IEEE Globecom Workshops
  (GC Wkshps)}.\hskip 1em plus 0.5em minus 0.4em\relax IEEE, 2021, pp. 1--6.

\bibitem{liu2023dhsa}
Z.~Liu, S.~Chen, J.~Ye, J.~Fan, H.~Li, and X.~Li, ``Dhsa: efficient doubly
  homomorphic secure aggregation for cross-silo federated learning,'' \emph{The
  Journal of Supercomputing}, vol.~79, no.~3, pp. 2819--2849, 2023.

\bibitem{fedmask}
A.~Li, J.~Sun, X.~Zeng, M.~Zhang, H.~Li, and Y.~Chen, ``Fedmask: Joint
  computation and communication-efficient personalized federated learning via
  heterogeneous masking,'' in \emph{Proceedings of the 19th ACM conference on
  embedded networked sensor systems}, 2021, pp. 42--55.

\end{thebibliography}
\appendices
\section{}
\label{app:A}

\begin{figure}[h!]
    \centering
    \includegraphics[clip, trim= 0cm 0cm 0cm 0.2cm, width=\linewidth]{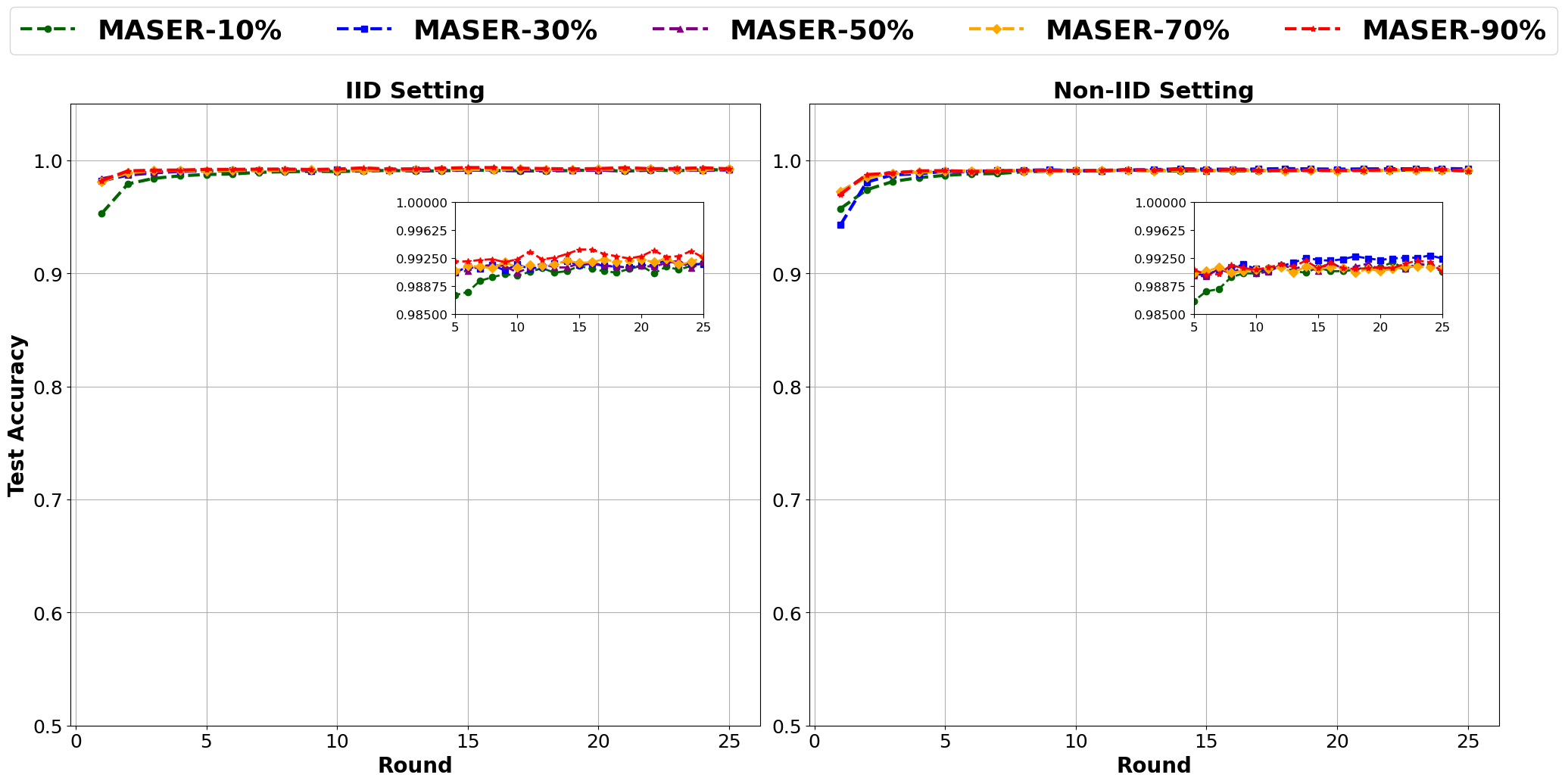}
    \caption{Test accuracy 
    across 25 FL rounds on MNIST with different thresholds}
    \label{fig:diff}
\end{figure}

Figure \ref{fig:diff} demonstrates the test accuracy of MASER across different pruning thresholds (i.e., different levels of sparsity) under both IID and Non-IID settings. Each curve corresponds to a different pruning level, where the percentage indicates the proportion of model weights retained after pruning. 
For instance, MASER-30\% keeps 30\% of the original weights with largest magnitudes and prunes the rest.
Despite the varying levels of sparsity, the accuracy curves remain close. This shows that our pruning strategy, combined with how we update important parameters, effectively preserves the important information needed for learning, even at high levels of sparsity.
The insets further highlight that although minor fluctuations exist, the convergence behavior and final accuracies are highly comparable.

\end{document}